\pdfoutput=1
\documentclass[journal,10pt]{IEEEtran}
\usepackage{latexsym}
\usepackage{graphicx}
\usepackage{amsfonts,amssymb,amsmath}
\usepackage{hyperref}
\def\BibTeX{{\rm B\kern-.05em{\sc i\kern-.025em b}\kern-.08em T\kern-.1667em\lower.7ex\hbox{E}\kern-.125emX}}

\usepackage{cite}
\newcommand{\ignore}[1]{}  % {} empty inside = %% comment
\usepackage{amsmath}

\usepackage[dvipsnames]{xcolor}
\usepackage{boxhandler}
\usepackage{siunitx}
\usepackage{balance}
\usepackage{cleveref}
  \usepackage{times}
\usepackage[T1]{fontenc}
\usepackage{mwe}    % loads »blindtext« and »graphicx«
\usepackage{subfigure}

\usepackage {optidef}
\usepackage{tablefootnote}

\usepackage{footnote}
\makesavenoteenv{tabular}
\usepackage[para,online,flushleft]{threeparttable}
\usepackage{enumitem}

\usepackage{booktabs, tabularx}

\usepackage[T1]{fontenc}
\usepackage{array}
\newcolumntype{P}[1]{>{\centering\arraybackslash}p{#1}}
\usepackage{multirow}
\usepackage{multicol}
\usepackage{mathtools}
\usepackage{graphicx}
\usepackage{balance}

\usepackage{float}

\begin{document}
\title{

%UAV Detection and Identification Using RF Fingerprints and Hierarchical Classification Framework

Hierarchical Learning Framework for UAV Detection and Identification
}

\author{%
Olusiji O Medaiyese, Martins Ezuma, Adrian P Lauf, and Ayodeji A Adeniran %
% \vspace{-5mm}
\thanks{This work has been supported in part by NASA under the Federal Award ID number NNX17AJ94A, and by NSF CNS-1939334 Aerial Experimentation Research Platform for Advanced Wireless (AERPAW) project that supported the experiments at NC State.}

% \thanks{All the authors are with the Department of Electrical and Computer Engineering, North Carolina State University, Raleigh, NC 27606 (e-mail:
% \{mcezuma, iguvenc\}@ncsu.edu).}%
}
% \maketitle
% \thispagestyle{plain}
% \pagestyle{plain}

\maketitle
\begin{abstract}

The ubiquity of unmanned aerial vehicles (UAVs) or drones is posing both security and safety risks to the public as UAVs are now used for cybercrimes. To mitigate these risks, it is important to have a system that can detect or identify the presence of an intruding UAV in a restricted environment. In this work, we propose a radio frequency (RF) based UAV detection and identification system by exploiting signals emanating from both the UAV and its flight controller, respectively. While several RF devices (i.e., Bluetooth and WiFi devices) operate in the same frequency band as UAVs, the proposed framework utilizes a semi-supervised learning approach for the detection of UAV or UAV's control signals in the presence of other wireless signals such as Bluetooth and WiFi. The semi-supervised learning approach uses stacked denoising autoencoder and local outlier factor algorithms. After the detection of UAV or UAV's control signals, the signal is decomposed by using Hilbert-Huang transform and wavelet packet transform to extract features from the time-frequency-energy domain of the signal. The extracted feature sets are used to train a three-level hierarchical classifier for identifying the type of signals (i.e., UAV or UAV control signal), UAV models, and flight mode of UAV. To demonstrate the feasibility of the proposed framework, we carried out an outdoor experiment for data collection using six UAVs, five Bluetooth devices, and two WiFi devices. The acquired data is called Cardinal RF (CardRF) dataset, and it is available for public use to foster UAV detection and identification research.
\end{abstract}


\IEEEpeerreviewmaketitle

\section{Introduction}

There is a steady growth in the application of unmanned aerial vehicles (UAVs) or drones across healthcare, agriculture \cite{alsalam2017autonomous}, environmental and disaster management \cite{coveney2017lightweight} and many other fields. For instance, technology companies are beginning to use UAVs for goods and service delivery. Recently, Amazon Inc's UAV delivery fleet received Federal Aviation Administration (FAA) clearance \cite{amazon2020FAA} to use UAVs for the delivery of goods and services. Other companies such as UPS Inc., and Wing, a subsidiary of Alphabet Inc., \cite{alphabet2020FAA} have also received their FAA clearance. The bid to integrate civilian UAVs into the national low altitude airspace by the FAA's Next Generation Air Transportation System (NextGen) and  National Aeronautics and Space Administration's (NASA) Unmanned Aircraft Systems Traffic Management project is already underway \cite{altawy2016security, NASA2020utm}. So, UAVs are becoming an integral part of our society, and more than 70 percent of the registered UAVs in the US are used for recreation activities \cite{uasbythenumbers_2020}.

However, there are increasing security, safety and privacy concerns in deploying civilian UAVs into the national airspace. Few of the menaces of UAVs include illegal surveillance or terrorist attacks. For instance, a civilian UAV crashed into the army's helicopter \cite{civilian2017army}. As a result of the malicious use of UAVs, there is a need for systems capable of detecting or identifying the presence of UAV in an environment or airspace. In \cite{altawy2016security}, access policies and UAV tracking are proposed as a defense to an intruding UAV. The access policies involve the creation of the No-Fly-Zone register or list that UAV manufacturers can incorporate in their UAV firmware. The pre-programming of the No-Fly-Zone list in a UAV firmware is called geofencing. UAVs with geofencing capabilities are restricted from flying into or taking off within the areas marked as No-Fly Zones based on global positioning system information. On the other hand, UAV tracking involves detecting and monitoring the presence of UAV in an environment or infrastructure and possibly alerting a watchman \cite{altawy2016security}. UAV tracking can be used in geofencing-free areas that are sensitive to security, safety, and privacy.

Some of the possible ways of detecting the presence of UAV in an environment include: RADAR, humming sounds from UAV (audio), video surveillance, thermal ray from UAV motor or rotors (thermal sensing), radio frequency (RF) signals from UAV controller or telemetry (RF sensing) or a combination of one or more sensing techniques (multimodality sensing). The comparison of these sensing techniques have been discussed in \cite{medaiyese2021wavelet}. RF sensing is adopted in this work because of its advantages over other forms of sensing techniques. Some of the advantages of using RF detection techniques are: i) its operation is stealthy; ii) it can detect UAVs of any size (i.e., mini, small, medium and so on); iii) it can detect UAVs that are in both line-of-sight and non-line-of-sight, and iv) it can be used for identifying the flight mode (i.e., flying, hovering, videoing, etc.) of UAVs \cite{al2019rf,alipour2019machine}.

The unmanned aerial system (UAS) is made up of a UAV, ground station (i.e., flight controller) and communication link between the UAV and its flight controller. Because the civilian UAS operates in the same industrial, scientific and medical (ISM) frequency band (i.e., 2.4 GHz) as Bluetooth and WiFi devices, it becomes more difficult to distinguish a UAS (i,e, UAV or UAV flight controller) signal from other ISM devices. This work attempts to provide a framework for detecting and identifying a UAV or UAV control signal in presence of Bluetooth and WiFi signals. For clarity, when we refer to a UAS signal it implies to either a UAV or UAV flight controller signal.

The contributions of this work are summarized as follows:\looseness=-1

\begin{enumerate}

\item We propose a framework for differentiating UAS signals from other wireless devices (i.e., Bluetooth and WiFi devices) operating at 2.4~GHz frequency. The framework uses a stacked denoising autoencoder (SDAE) for signal compression. A denoising autoencoder (DAE) is adopted to make our system robust to variation in signal to noise ratio (SNR) as a result of environmental factors or channel effects. The essence of signal compression is to reduce the space complexity of detecting any form of UAV in an environment. The compressed signal is a latent representation of a signal and it is passed into a local outlier factor (LOF) algorithm to detect UAS signals as anomalous signals from WiFi and Bluetooth signals. We increased the diversity of detecting a UAS by using signals emitting from both the UAV flight controller (i.e., control signals) and the airborne vehicle (i.e., telemetry and other secondary signals from the UAV).

\item When a UAS signal is detected, further classification of the UAS signal is carried out to determine the type of UAS signal (UAV or UAV controller signal), type of UAS model, and flight mode of the UAV. The Hilbert-Huang transform (HHT) and wavelet packet transform (WPT) are used to extract unique signatures or features from the steady state of the UAS signals. This work is a shift from the norm of extracting unique signature from the transient state of RF signals for device or UAV identification. The extracted feature set is then used to train a three level hierarchical classifier which uses extreme gradient boosting classifiers (XGBoost) as the learning algorithm.

\item We demonstrate the feasibility of the proposed framework by collecting UAS, Bluetooth and WiFi RF signals over-the-air in an outdoor environment. The acquired data is called the Cardinal RF (CardRF) dataset and it is made available for public use to foster research in UAV or rogue RF device detection and identification.
\end{enumerate}

The remainder of the paper is organized as follows:  Section~\ref{two} provides a brief overview of the related work. Section~\ref{three} describes the system model overview. Section~\ref{four} introduces the experimental setup and data capturing steps. Section~\ref{five} describes the proposed UAS detection framework using the SDAE and LOF algorithms. Section~\ref{six} introduces the UAV classification or identification framework; we discuss the feature extraction approaches and the proposed hierarchical classifier. In section~\ref{seven}, the performance evaluation and results are discussed. We provided the conclusion and future work of this paper in Section~\ref{eight}.

\section{Related Work} \label{two}
In the literature, different approaches have been proposed for RF-based UAV detection and identification systems. In \cite{zhao2018classification}, the authors proposed an Auxiliary Classifier Wasserstein
Generative Adversarial Networks (AC-WGANs) for UAV detection. Fewer UAVs' signals, WiFi signals, and random signals (i.e., unknown signals from the environment) with high dimensionality of 100,000 signal length are utilized. Amplitude envelope and principal component analysis (PCA) are used to reduce the dimensionality of each signal captured to 400. The authors leveraged the compressed signals to train AC-WGANs for UAV classification and an accuracy of 95\% was achieved at 5~dB and above. The signal length is high and it will increase the detection time. Minimal signal lengths can be used and also there is no results on average computational inference time for the AC-WGAN. We are also not sure about the performance of the approach for identifying UAV flight mode.

Because RF sensing techniques can be exploited for detecting the presence of UAVs, type of UAVs and the flight mode of UAVs, the authors in \cite{nemer2021rf} proposed a hierarchical classification approach to RF-based UAV detection and identification system. An accuracy of 99.2\% is achieved for detecting the operation modes of UAVs. However, the authors use a flat classification metric for the evaluation and inference from other ISM devices is not considered. Similarly, an accuracy of 91\% was achieved in \cite{swinney2021unmanned} for flight mode identification of UAVs. The power spectral density (PSD) images of UAV signal are adopted as signatures. A deep residual learning (ResNet50) was used to extract features from the PSD images. The extracted features are used to train a logistic regression algorithm for classification. However, the authors in \cite{swinney2021unmanned} did not evaluate their proposed models under varying SNR, and the discrimination for other ISM band signals was also not considered. More so, the computational inference cost of the proposed models in \cite{swinney2021unmanned} was not considered.

In \cite{medaiyese2021wavelet}, wavelet scattering transforms were used to extract image-based features (i.e., scattergram) from the UAV's control signals and a convolutional neural network algorithm, SqueezeNet was trained with the extracted features. An accuracy of 98.9\% was achieved at 10~dB SNR. However, the authors only use the UAV control signals for UAV detection. Similarly, in \cite{ezuma2019detection,medaiyese2021semi}, the transient state of the UAV control signals is exploited for UAV detection. The transient state RF signal are significantly affected by channel effect \cite{medaiyese2021wavelet}. So, it is difficult to detect or capture the transient state at a very low SNR. More so, relying only on UAV control signals could be limited because the UAV can be remotely and intermittently controlled from several miles away.

An algorithm for classifying identical UAVs (i.e., same make and model) in hovering flight mode was proposed in \cite{soltani2020rf}. Although an overall accuracy of 91\% was achieved, the authors did not consider interference mitigation strategy for other ISM devices operating at 2.4~GHz. Also, the use of multiple neural networks (NN) as classifiers increases the complexity of their proposed algorithm and the only flight mode considered was hovering.

As seen in \cite{medaiyese2021wavelet}, when building a machine learning model for an RF-based UAV detection and identification system, having high classification accuracy does not connote that the model is robust to variation in wireless channel. For instance, building and evaluating a model using training and test sets with the same SNR. In reality, the system should be resilient to variation in SNR due to changes in environmental conditions or other channel effects.

To the best of our knowledge, there are only four publicly-available dataset for RF-based UAV detection research. DroneRF dataset was proposed in \cite{al2019rf} and it has been actively used in the UAV research space. Examples of papers where this dataset have been used are: \cite{akter2020rf, al2020drone, nemer2021rf,swinney2020unmanned,swinney2021unmanned}. However, avenues to enable the research community to factor interference from commonplace signals (i.e., Bluetooth and WiFi signals) in their RF-based UAV detection system design is not possible using this dataset. Similarly, a drone radio controller RF dataset was proposed in \cite{martins2020controller}, but it strictly contains only control signals from UAV flight controllers. As observed in our experiment in Section \ref{four}, if the UAV is closer to the detection system than the flight controller, the detection system only captures the non-control signals from the UAV. In this case, using only the control signals for UAV detection reduces the detection probability and it is difficult to exploit the control signals for UAV flight mode identification. Furthermore, DroneSignal was proposed in \cite{basak2021drone} which includes UAV controls and video signals and one WiFi signal. However, the majority of the UAV signals are control signals. The more recent dataset is the hovering UAVs RF fingerprinting dataset \cite{soltani2020rf}, where seven identical (i.e., same make and model) UAVs were utilized. However, the only flight mode considered is the hovering mode. Hence, we propose the CardRF to fill some of the loopholes in other datasets.

\section{System Model Overview} \label{three}
\begin{figure}[h]
\center{\includegraphics[scale=0.26]{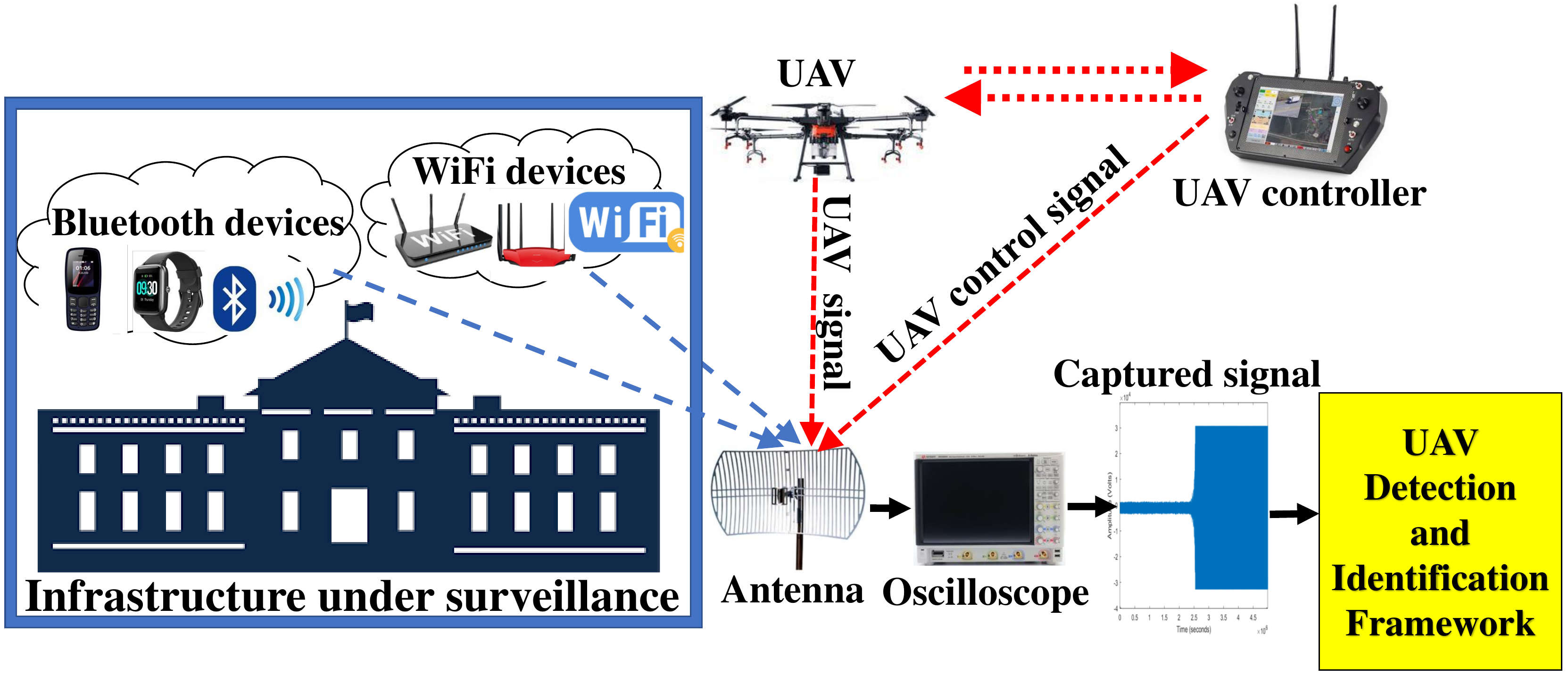}}
\caption{ A scenario of RF-based UAV detection and Identification system as applicable for infrastructure surveillance in the presence of other ISM devices (WiFi and Bluetooth devices). Both the UAV controller signal and the signal from UAV are exploited for UAV detection and identification.}
\label{Fig:system_model}
 \vspace{-2mm}
\end{figure}
The system modeling of an RF-based UAV detection and identification system and how it can be used to monitor an infrastructure is illustrated in Fig~\ref{Fig:system_model}. Here, it is assumed that recognized signals from Bluetooth and WiFi signals are permitted to propagate within or around the infrastructure. On the contrary, no UAV is allowed to take off or fly into the the infrastructure vicinity. More so, it is assumed that the UAV and its flight controller operate at the same 2.4~GHz frequency band as the Bluetooth and WiFi device. A detection system can be set up to passively detect and analyze any signal propagating at 2.4~GHz in order to identify UAS signals.

\section{Experimental Setup and Data Capturing}\label{four}
\begin{figure}[h]
\center{\includegraphics[clip,scale=0.45]{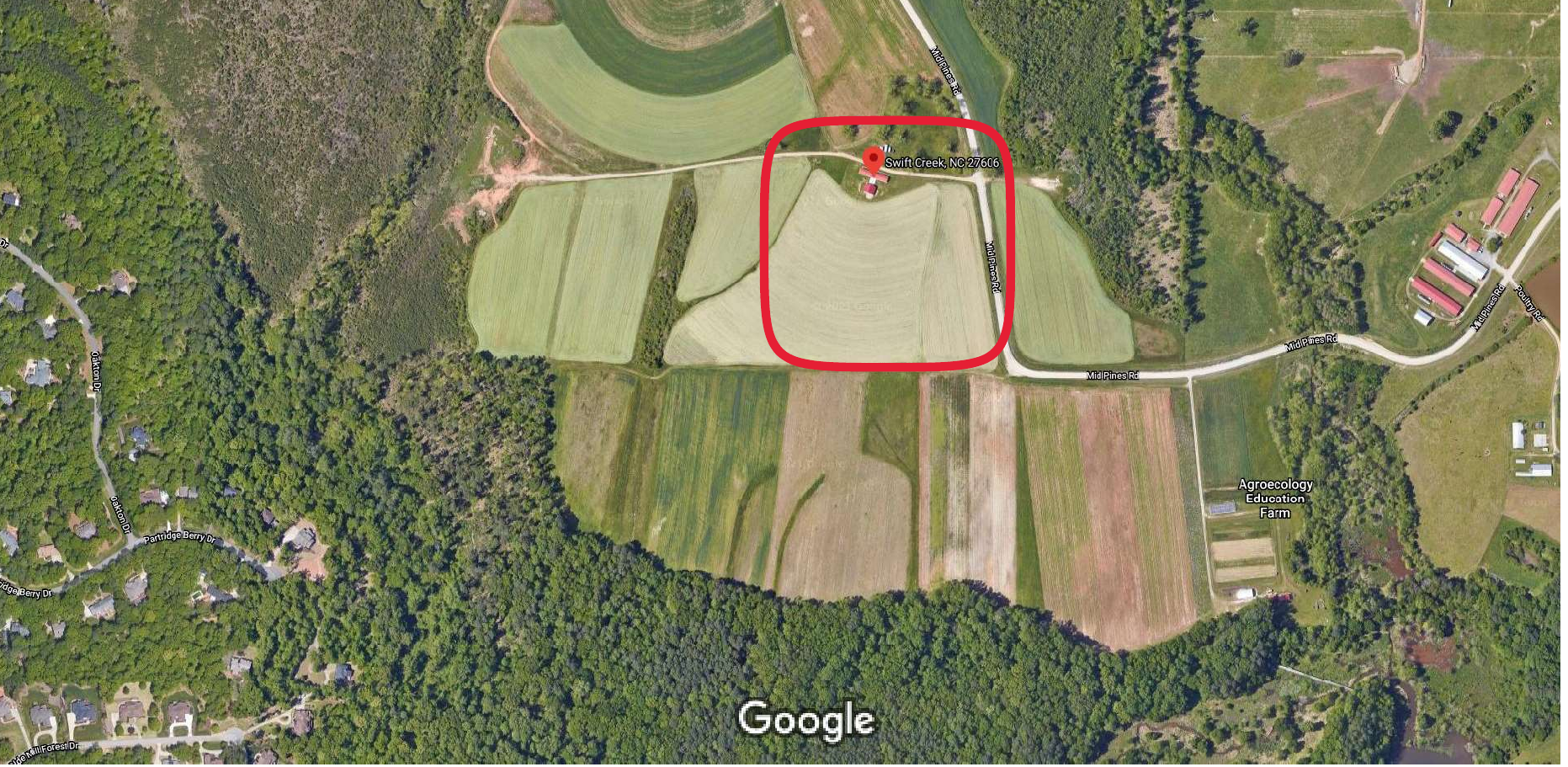}}
\caption{ The satellite view of the experiment location from Google map. The location of experiment is 4191 Mid Pines road, Raleigh, North Carolina, USA. }
\label{Fig:experiment_site_map}
 \vspace{-5mm}
\end{figure}

In this section, we describe the CardRF dataset. The major reasons for proposing this dataset are:
\begin{enumerate}
    \item To give the research community the avenue to build a UAV detection system that can detect UAVs in the presence of other wireless interference such as Bluetooth or/and WiFi devices.  
    \item Opportunity to use both the UAV's control and non-control (i.e., telemetry and video) signals for UAV detection system and evaluate the system under varying SNR.
\end{enumerate}
The device catalogue (UAS, Bluetooth, and WiFi) used for the data acquisition is listed in Table \ref{device_catalogue}. The Bluetooth and WiFi devices are used to acquire Bluetooth and WiFi signals, respectively. Similarly, two components (i.e., UAV and its flight controller) of a UAS are utilized in collecting the UAS signals.
\begin{table}[h]
\centering
\caption{ Catalog of RF devices used in the experiment for 2.4~GHz RF fingerprint acquisition.}
\label{device_catalogue}
\begin{tabular}{|c|c|c|}
\hline
Device & Make & Model\\
\hline
UAV &\multirow{4}{*}{\text{ DJI}} & Phantom 4  \\
&&Inspire  \\
&&Matrice 600  \\
&&Mavic Pro 1  \\\cline{2-3}
& Beebeerun & FPV RC drone mini quadcopter\\\cline{2-3}
& 3DR & Iris FS-TH9x\\
\hline
\multirow{2}{*}{\text{Bluetooth }}& \multirow{3}{*}{\text{Apple}} & iPhone 6S  \\
&&iPhone 7  \\
&&iPad 3  \\\cline{2-3}
& FitBit  &  Charge3 smartwatch\\\cline{2-3}
& Motorola & E5 Cruise \\
\hline
\multirow{2}{*}{\text{WiFI}} &Cisco & Linksys E3200\\\cline{2-3}
&TP-link & TL-WR940N\\
\hline
\end{tabular}
% \vspace{-3mm}
\end{table}
The outdoor experiment was conducted at the Lake Wheeler site located at 4191 Mid Pines road, Raleigh, North Carolina, USA. Fig.~\ref{Fig:experiment_site_map} shows the satellite view of the location where the experiment was carried out. It should be noted that the experiment was conducted under a FAA special airworthiness certificate and that the UAVs were operated by FAA certified pilots.

\subsection{Experimental Design}
\begin{figure*}[t]
\center{
\begin{subfigure}[]{\includegraphics[scale=0.22]{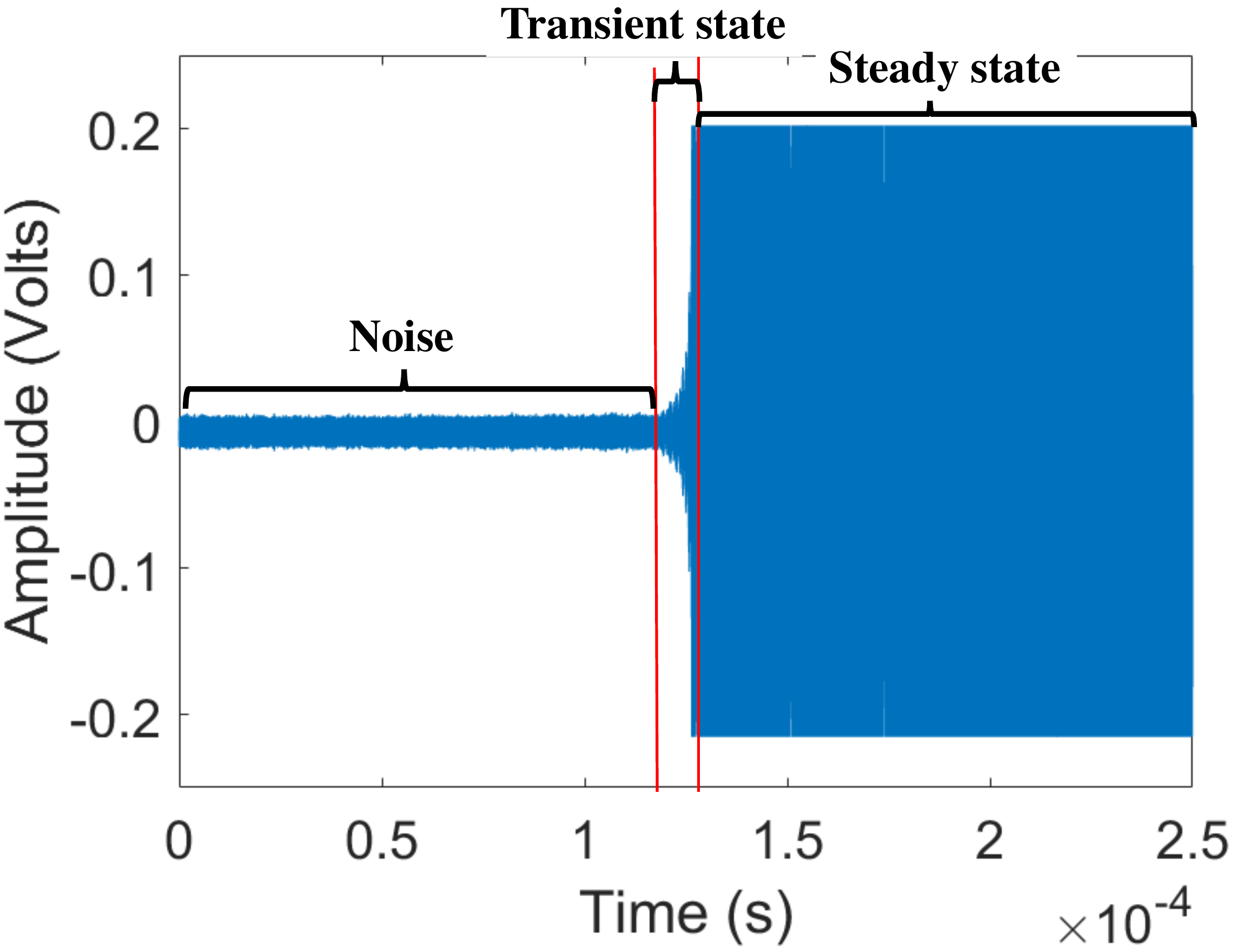}\label{}}
\end{subfigure}
\hspace{2mm}
\begin{subfigure}[]{\includegraphics[scale=0.25]{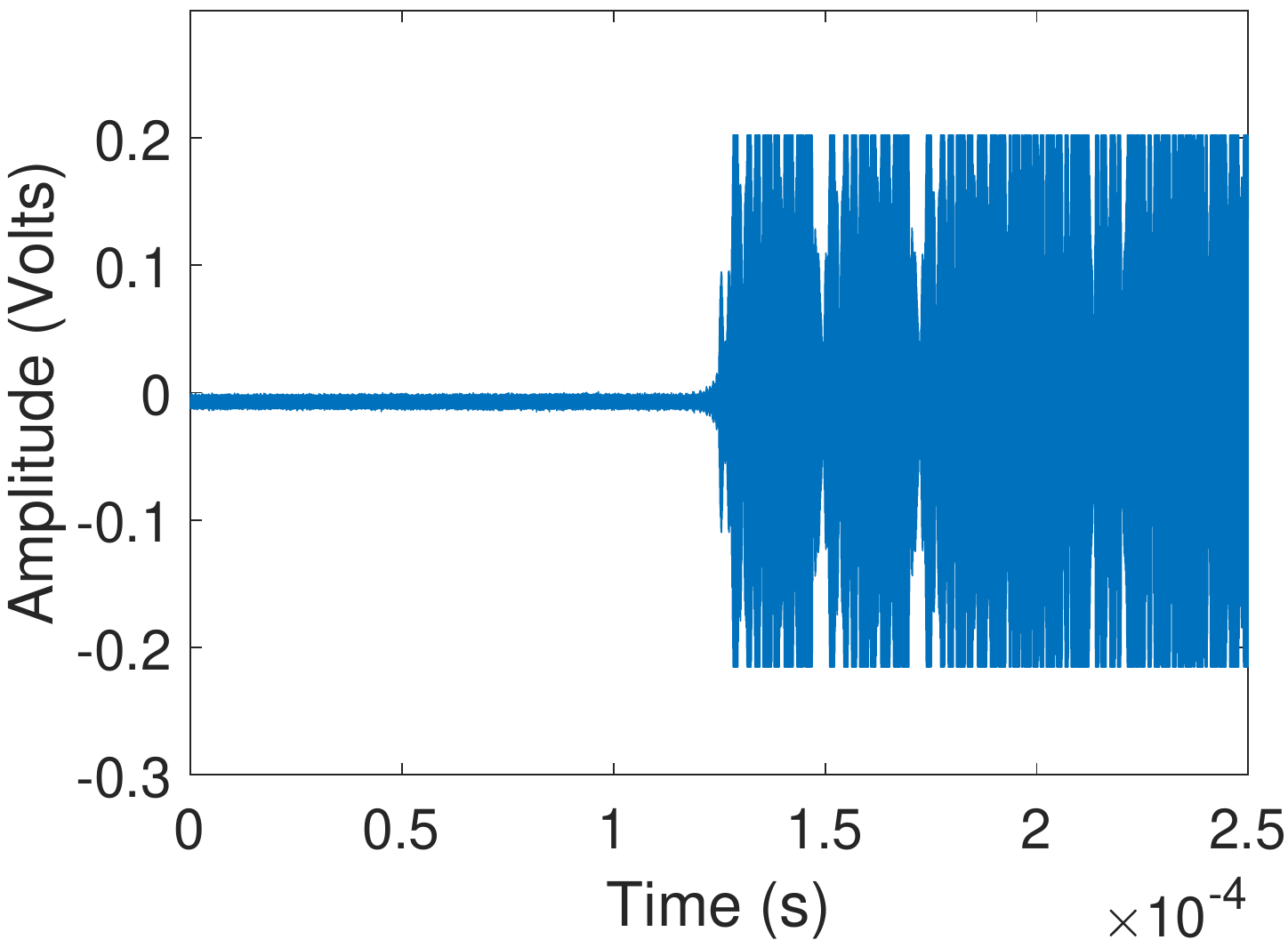}\label{}}
\end{subfigure}
\hspace{2mm}
\begin{subfigure}[]{\includegraphics[scale=0.25]{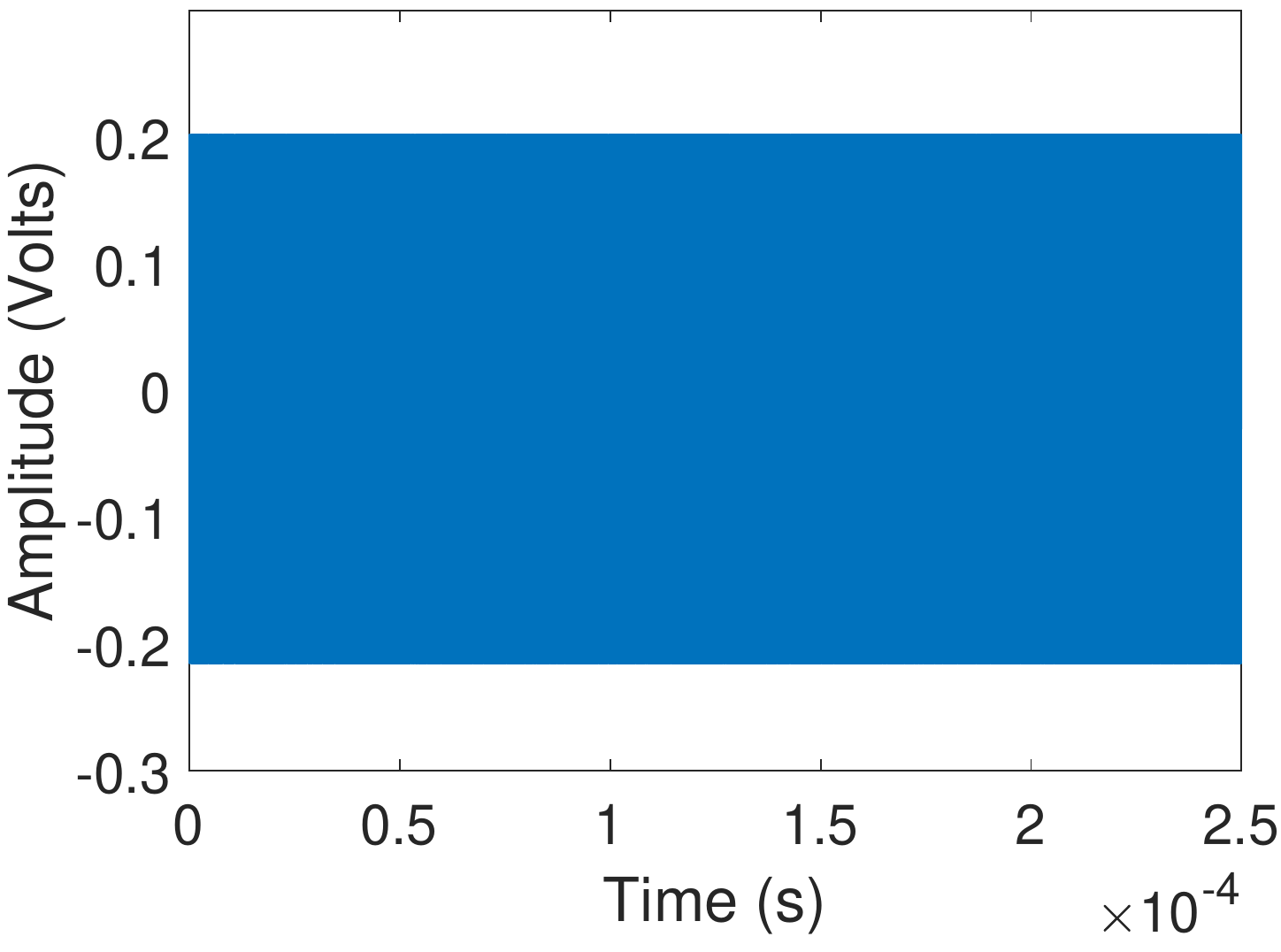}\label{}}
\end{subfigure}
\hspace{2mm}
\begin{subfigure}[]{\includegraphics[scale=0.25]{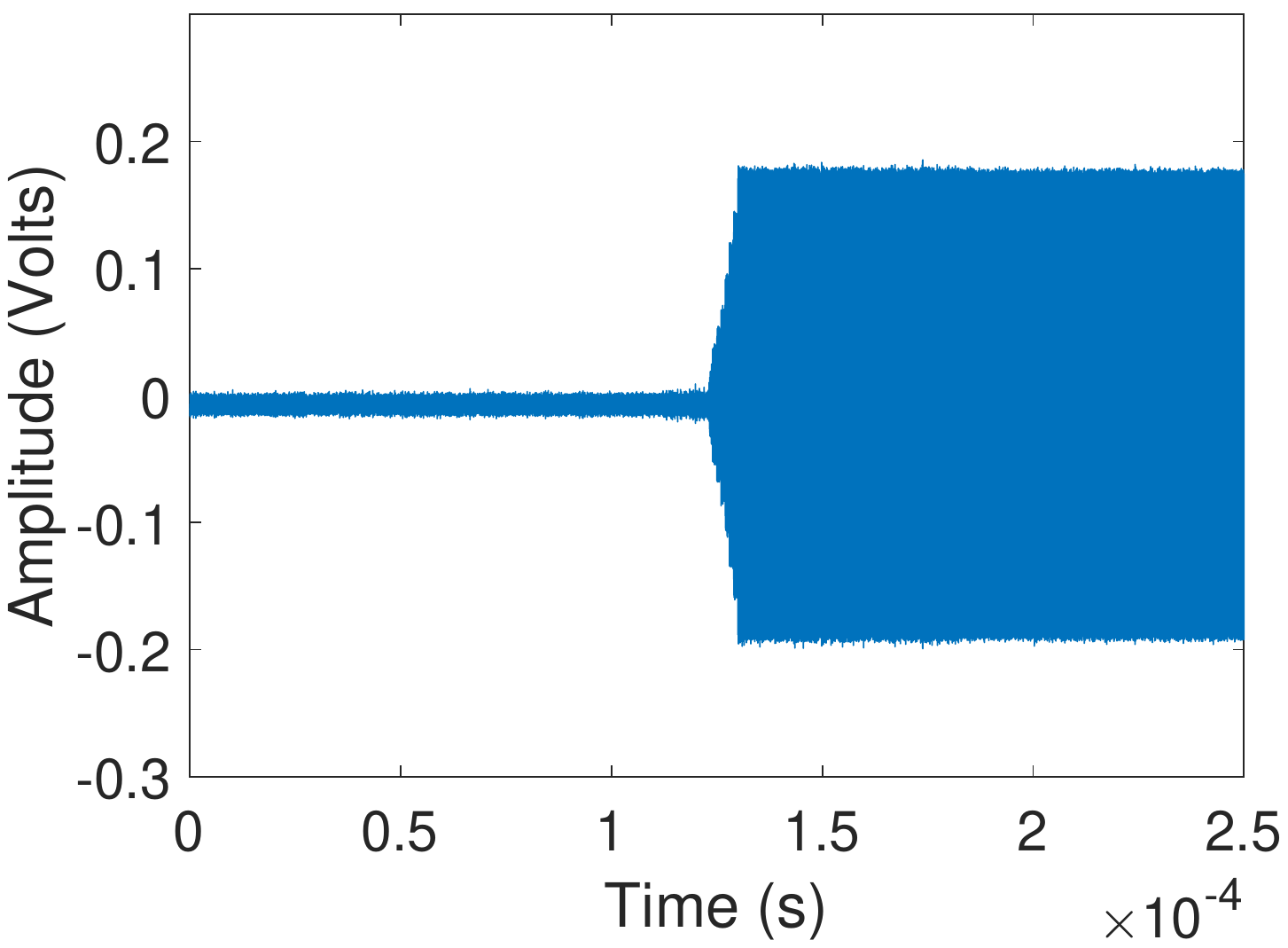}\label{}}
\end{subfigure}
\hspace{2mm}
 \caption{{Captured RF signal from: (a) DJI Phantom 4, (b) DJI Matrice 600 controller, (c) DJI Matrice 600 UAV (d) Beebeerun controller. In (a) the RF signal from DJI Phantom 4 controller with various parts (i.e., transient and steady state) of the signal are labeled.  }}
  \label{signal_from_devices}}
  \vspace{-3mm}
\end{figure*}
The electronic system for the data acquisition consists of five major components. These components include: a 24 dBi 2.4 GHz grid parabolic antenna, a 2.4 GHz bandpass filter, an RF low noise amplifier (LNA), a DC (direct current) generator, and an oscilloscope. The directional antenna is the front-end component of the detection system and it is used to pick up any electromagnetic waves (i.e., RF signals) propagating at a 2.4~GHz frequency. To ensure that only 2.4~GHz signals are acquired, an Airvu 2.4 GHz bandpass filter is connected to the antenna. The bandpass signal is amplified by using a LNA, FMAM63007 which operates from 2~GHz to 2.6~GHz with 30~dB gain. The LNA is powered by using a DC generator. The amplified signal is sampled at 20~GSa/s by the 6 GHz bandwidth Keysight MSOS604A oscilloscope.

\begin{figure}[h]
\center{\includegraphics[scale=0.28]{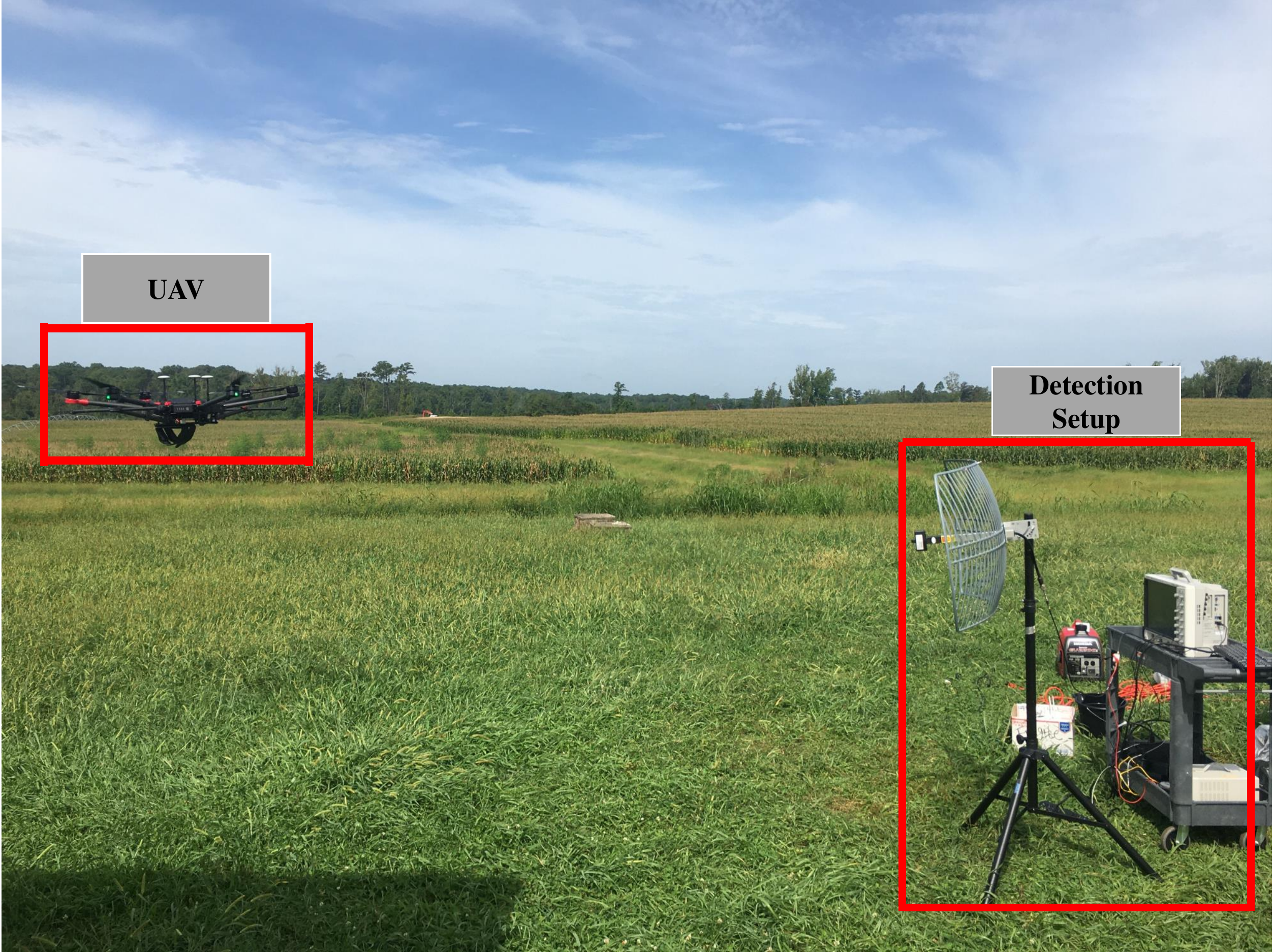}}
\caption{The outdoor experimental setup showing the signal capturing system and a UAV. }
\label{Fig:experiment_setup}
 \vspace{-5mm}
\end{figure}

Before capturing signals from the devices, it is important to consider the background noise in the area. The oscilloscope was calibrated to detect signals above the background noise by using the threshold trigger functionality. The background noise level in the environment was observed in the absence of RF signals and the energy level is set as the threshold. When the energy level is above the threshold it implies the presence of a signal. The oscilloscope captures data when the energy level is greater than the threshold and goes into idle mode when the energy level is equal to or less than the threshold. The later connotes the absence of signal in an environment.
The signals were captured at a distance range between 8 - 12~meters from the detection system. The signals from UAVs were captured under different flight modes (i.e., flying, videoing and hovering). Fig.~\ref{Fig:experiment_setup} shows the experimental setup when capturing a UAV signal.
\subsection{Description of Data}

The signals are sampled and each of the signal consists of five million sampling points. Fig~\ref{signal_from_devices}(a) shows an RF signal captured from a DJI Phantom 4 flight controller where the transient and steady state of the signals are labeled. Similarly,  Fig~\ref{signal_from_devices} (b)-(d) shows the typical examples of signals captured from DJI Matrice 600 controller, DJI Matrice 600 UAV, and Beebeerun flight controller, respectively. It can be easily seen that the transient state appears to be missing in the signal captured from DJI Matrice 600. It was also observed that majority of the non-control signals from DJI UAVs do not have a transient state. So exploiting the signal's transient state as fingerprint might not always be reliable for UAV detection. For this reason, the steady state of RF signals is utilized. More so, it has been shown in the literature that channel noise significantly affect transient state more than steady state of RF signal when using RF signals for device identification.

For the essence of this work, we sliced the steady state of the captured signals into 1024 sampling points per slice. The motivation for using 1024 sampling points per slice is to use minimum signal length. Using minimal signal length will enhance the time complexity of the UAV detection and identification system.

\section{UAS Detection Framework}\label{five}
The proposed framework is a multistage framework and it is divided into two stages. These include:
\begin{enumerate}
    \item Classification of UAS signals from interference signals like Bluetooth and WiFi signals. 
    \item Classification of UAV signals, model types, and UAV flight mode.
\end{enumerate}
In this section, we discuss the first level or stage of the framework. The classification of UAS signals from the interference signals is done by using the principles of anomaly detection approach. The SDAE-LOF model is used as the anomaly detector.
\subsection{Stacked Denoising Autoencoder}
Data explosion is one of the key challenges of using RF sensing for UAV detection especially when performing spectral analysis on the captured RF signal. Dimensionality reduction is a pathway to minimizing the explosion of data during analysis or inference \cite{kantardzic2011data}. This can be viewed as signal compression where a raw signal is compressed or transformed into a latent space representation. There are numerous dimension reduction algorithms (e.g., principal component analysis) and the choice of dimension reduction algorithm determines the performance quality. An autoencoder (AE) is an unsupervised neural network algorithm used for data compression, decompression and feature extraction. Encoding, code and decoding layers are the three essential parts of AE. The main goal of AE is to learn an efficient way to encode its input into a lower dimensional space and use the encoded respresentation (i.e., code) to reconstruct the original input data as the output. DAE is a variant of AE which enhances the ability to learn non-linear latent representation of a signal with robustness to signal corruption \cite{vincent2010stacked}.

\begin{figure}
\center{\includegraphics[clip,scale=0.28]{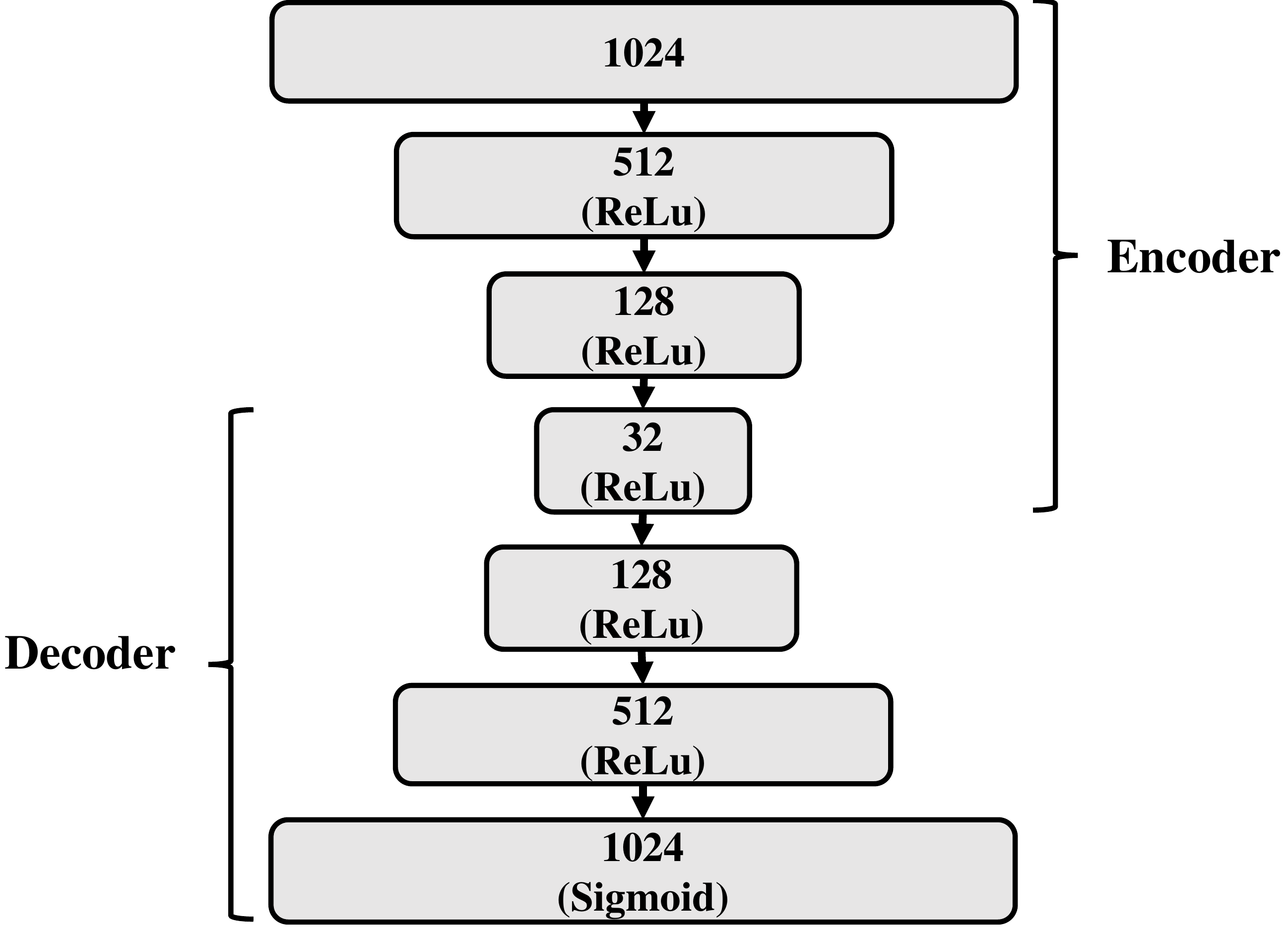}}
\caption{Architecture of SDAE for signal compression. The input and output layer have the same size (i.e., 1024) and the code layer is 32 dimensional size. The SDAE compress signal into a latent representation that is 32 dimensions. All the hidden layer uses ReLu as the activation function and the output layer utilizes the sigmoid as the activation function. }
\label{Fig:sdae_architecture}
 \vspace{-2mm}
\end{figure}

\begin{figure}[h]
\center{\includegraphics[ clip,scale=0.41]{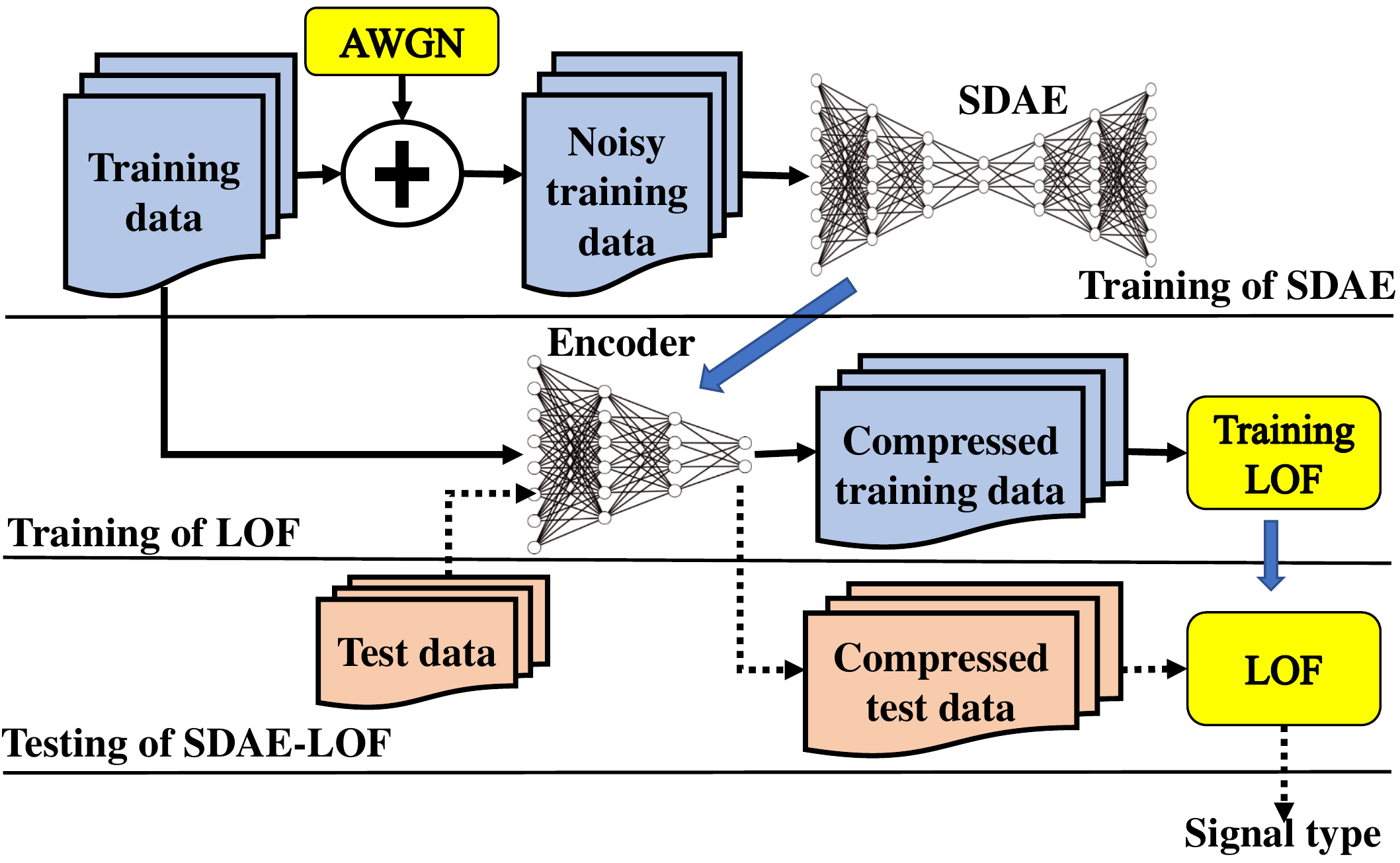}}
\caption{ Stacked denoising autoencoder-local outlier factor (SDAE-LOF) training process for UAV Detection. The training process is a dual level process. The first level deals with training the SDAE and the second level deals with training the LOF. After the training of SDAE, the encoder is detached from the network and used in the second stage to compress the uncorrupted training data. The compressed training data is used to training the LOF model. Similarly, the testing data is first compressed using the encoder before passing to the trained LOF for UAV detection.}
\label{Fig:sdae_model_pipeline}
\vspace{-3mm}
\end{figure}

We propose a SDAE for quick signal compression with robustness to channel noise. The architecture of the SDAE is shown in Fig.~\ref{Fig:sdae_architecture}. The input layer to third hidden layer represents the encoder. The third hidden layer represents the code layer and the third hidden layer to the output layer represents the decoder. The dimension of the input and output is 1024 and the code layer is 32. This implies that we use the SDAE to compress the signal from 1024 to 32 sampling points.
After the training process, only the encoder is used to compress the signals. Fig.~\ref{Fig:sdae_model_pipeline} illustrates the training process of the SDAE-LOF for detection of UAVs. 
The commonplace signal $x$ in the training set $T$ is assumed to be a recognized signal propagating within the infrastructure under surveillance at a high SNR where there is little or no channel noise. Signal $x$ is corrupted as $\acute{x}$ due to channel noise. $\acute{x}$ is given as:
\begin {equation} \label{eq:1}
\acute{x}= x + n,
\end{equation}
where $n$ is the additive white Gaussian noise (AWGN). We assume that the channel noise is AWGN in this work.

Each signal $x$ in $T$ is randomly corrupted with AWGN. The essence of the SDAE is to map  $\acute{x}$ to $x$ by forming a latent representation. To ensure a fast learning process by the SDAE, min-max normalization is applied on  $\acute{x}$ and $x$ in $T$ which constrained the signal amplitude between $0$ and $1$. The normalization of $\acute{x}$ and $x$ is given as $\acute{m}$ and $m$, respectively.

\begin {equation} \label{eq:2}
\acute{m}= \frac{{\acute{x}_i}- min (\acute{x}_{train})}{max (\acute{x}_{train})-min (\acute{x}_{train})},
\end{equation}

Normalized signals $\acute{m}$ and $m$ is mapped into a latent representation $h$ as: 
\begin {equation} \label{eq:3}
h= H(\acute{m}),
\end{equation}
where $H(.)$ is the encoder function of the SDAE.
 
By using $h$ to reconstruct $\acute{m}$ through a decoder function as $u$; $u$ must be nearly the same as $m$. So SDAE is trained to minimize the reconstruction error between $u$ and $m$.
The encoding  $H(\acute{m})$ is used to encode the captured signal into a latent representation, $h$,  which is fed into the detection algorithm for the detection of anomalies.

\subsection{Local Outlier Factor} 
LOF is an approach of assigning an object in a dataset a degree of being an outlier by measuring the local distance from its surrounding neighborhood \cite{breunig2000lof}. To illustrate an outlier using clustering technique, an outlier is an object that is isolated from clusters. Local and global outlier are two possible types of outlier. 
\begin{figure}[h]
\center{\includegraphics[clip,scale=0.4]{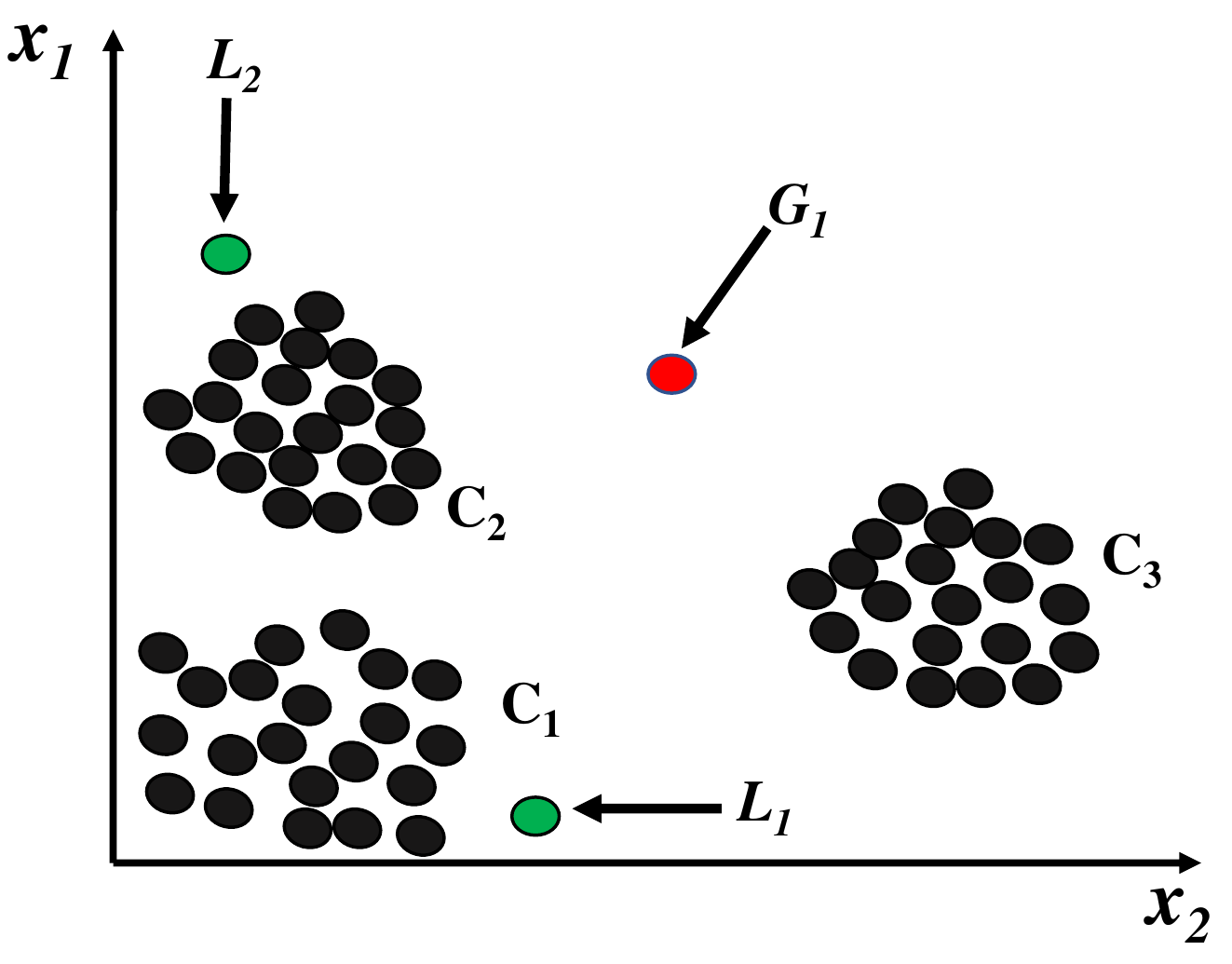}}
\caption{Graphical illustration of local and global outliers in a three cluster dataset (adopted from \cite{medaiyese2021semi}).}
\label{Fig:lof}
 \vspace{-2mm}
\end{figure}

Fig.~\ref{Fig:lof} shows a graphical representation of two dimensional dataset with three clusters (i.e., $C_1, C_2$ and $C_3$) and outliers (i.e., $L_1, L_2$ and $G_1$). $L_1$ and $L_2$ are local outliers to cluster $C_1, $ and $C_2$, respectively. On the other hand, $G_1$ is a global outlier to the three clusters. The problem of using a global outlier detection algorithm for this type of dataset with three cluster is that it might not detect $L_1$ and $L_2$ as an outlier. However, a local outlier detection algorithm would detect $L_1$ and $L_2$  and even detect $G_1$ as outlier. The LOF principles require four steps and these steps are briefly summarized as follow \cite{breunig2000lof}:
\begin{enumerate}
    \item Estimate the $k$-distance for the $k$th nearest neighbor. For instance, if $k$ is 5 then $k$-distance is the distance of a point to the fifth nearest neighbor. Given object $p$, the distance of object $p$ to $k$th nearest neighbor is denoted as $k$-distance$(p)$
    \item Estimate the reachability distance using the $k$-distance. The reachability distance is calculate as:
        \begin{equation}
        {reach\text{-}dist}_k(p,o)= max\{k\text{-}distance(o), d(p,o) \}.
        \end{equation}
    This implies that if data point $p$ within the $k$th nearest neighbor then the reachability distance (p,o) will be equal to the $k$-distance$(o)$, else, it will be the actual distance from $o$ to $p$ (i.e., $d(p,o)$). 
    \item Calculate the local reachability density (lrd) using the reachability distance. To estimate the lrd for a given point $p$, first, the reachability distances of $p$ to all its $k$ nearest neighbors are estimated. Secondly, the average value of all the calculated reachability distances of $p$ to all its $k$ nearest neighbors is estimated. The inverse of the average is the local reachability density. The lrd is mathematically expressed as:
        \begin{equation}
        lrd(p)=\frac{1}{ \Bigg( {\frac{\sum_{o \in N_{k}}reach\text{-}dist (p,o)}{k}} \Bigg)}.
        \end{equation}
    where $N_k$ represents the set of $k$ nearest neighbors. 
    \item Estimate the LOF of the data point using lrd. Given that the point is $p$, the LOF of point $p$ is average ratio of p's $k$-neighbors' lrds to p's lrd. It is expressed as:
        \begin{equation}
        LOF(p)= {\frac{\sum_{o \in N_{k}} \frac{lrd(o)}{lrd(p)}}{k}}.
        \end{equation}
            \begin{itemize}
             \item if $LOF(p) \approx 1$, it means $p$ has a similar density as its neighbors (i.e., not an outlier),
             \item if $LOF(p) < 1$, it means $p$ has a higher density than its neighbors (i.e., not an outlier),
             \item if $LOF(p) > 1$, it means $p$ has a lower density than its neighbors (i.e., an outlier).
            \end{itemize}
\end{enumerate}
The most important hyperparameters of LOF are the number of nearest neighbors that defines the local neighborhood of an object and the distance metric \cite{breunig2000lof}.

\section{UAS Classification System} \label{six}
\subsection{HHT For Time-Frequency Domain Feature Extraction}
Extracting feature as fingerprint from RF signals is not a trivial task. Mostly time-frequency domain analytics are exploited for extracting intrinsic and hidden characteristics of a signal. HHT has been used as a time-frequency-energy domain analytics tool to extract hidden features or characteristics in signals \cite{ali2019assessment}. Empirical mode decomposition (EMD) and Hilbert transforms (HT) are both used to obtain HHT.
\begin{figure*}
\center{
\begin{subfigure}[]{\includegraphics[scale=0.50]{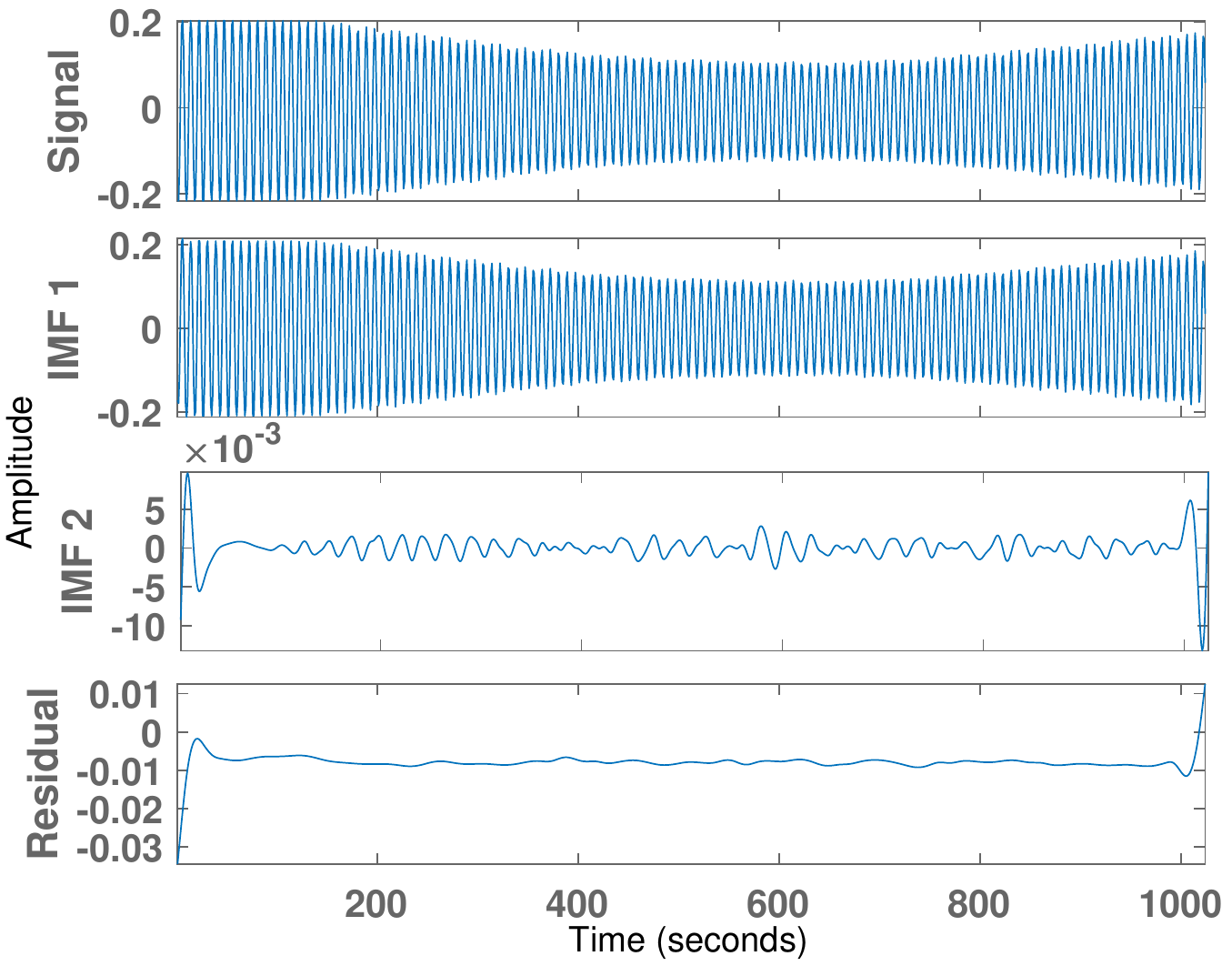}\label{}}
\end{subfigure}
\hspace{2mm}
\begin{subfigure}[]{\includegraphics[scale=0.50]{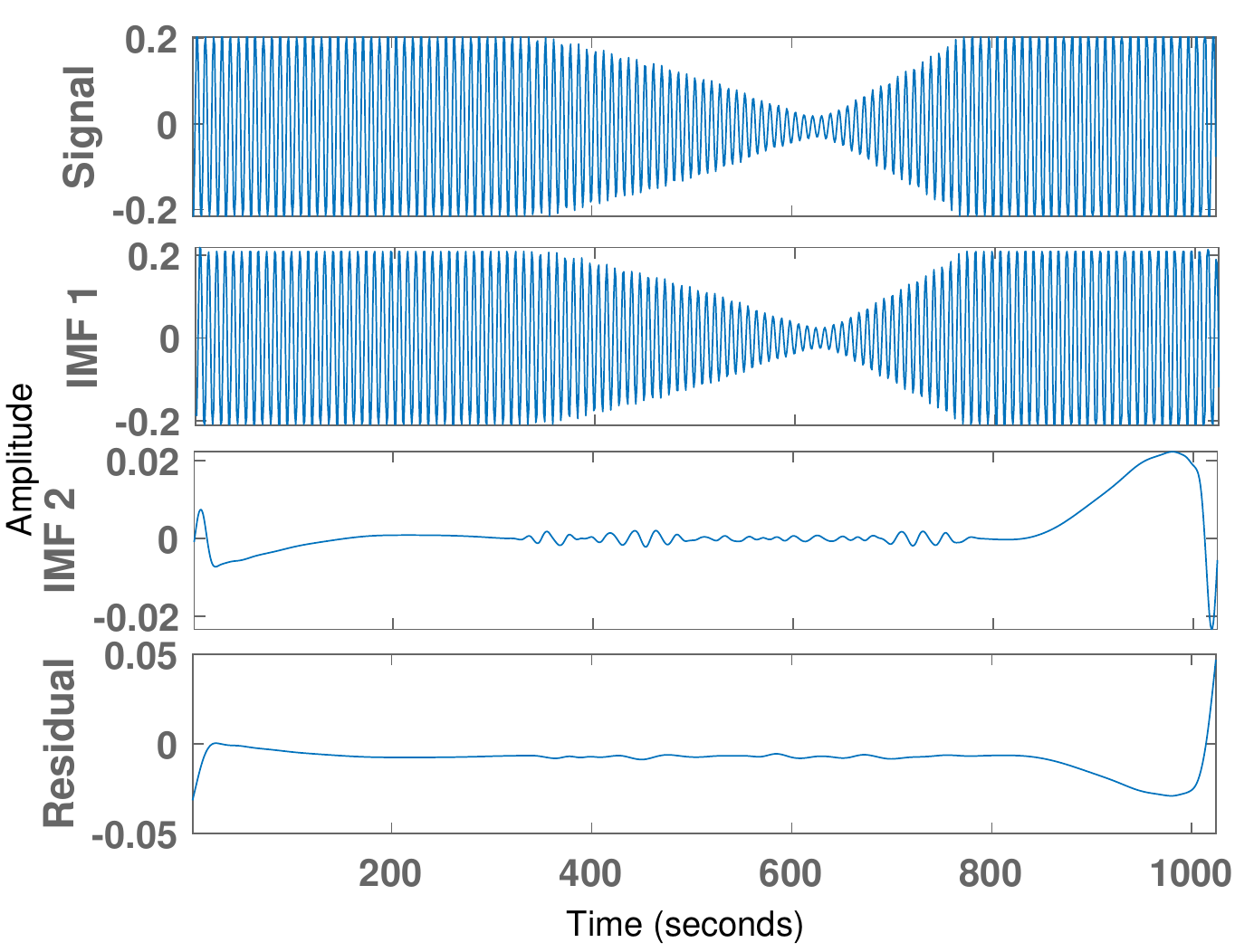}\label{}}
\end{subfigure}
 \caption{{The empirical mode decomposition showing the first two IMFs and the residual of the captured RF signals (i.e., using the steady state) from: (a) DJI Inspire, (b) Beebeerun.  }}
  \label{emd_hht}}
  \vspace{-5mm}
\end{figure*}
\begin{figure*}
\center{
\begin{subfigure}[]{\includegraphics[scale=0.48]{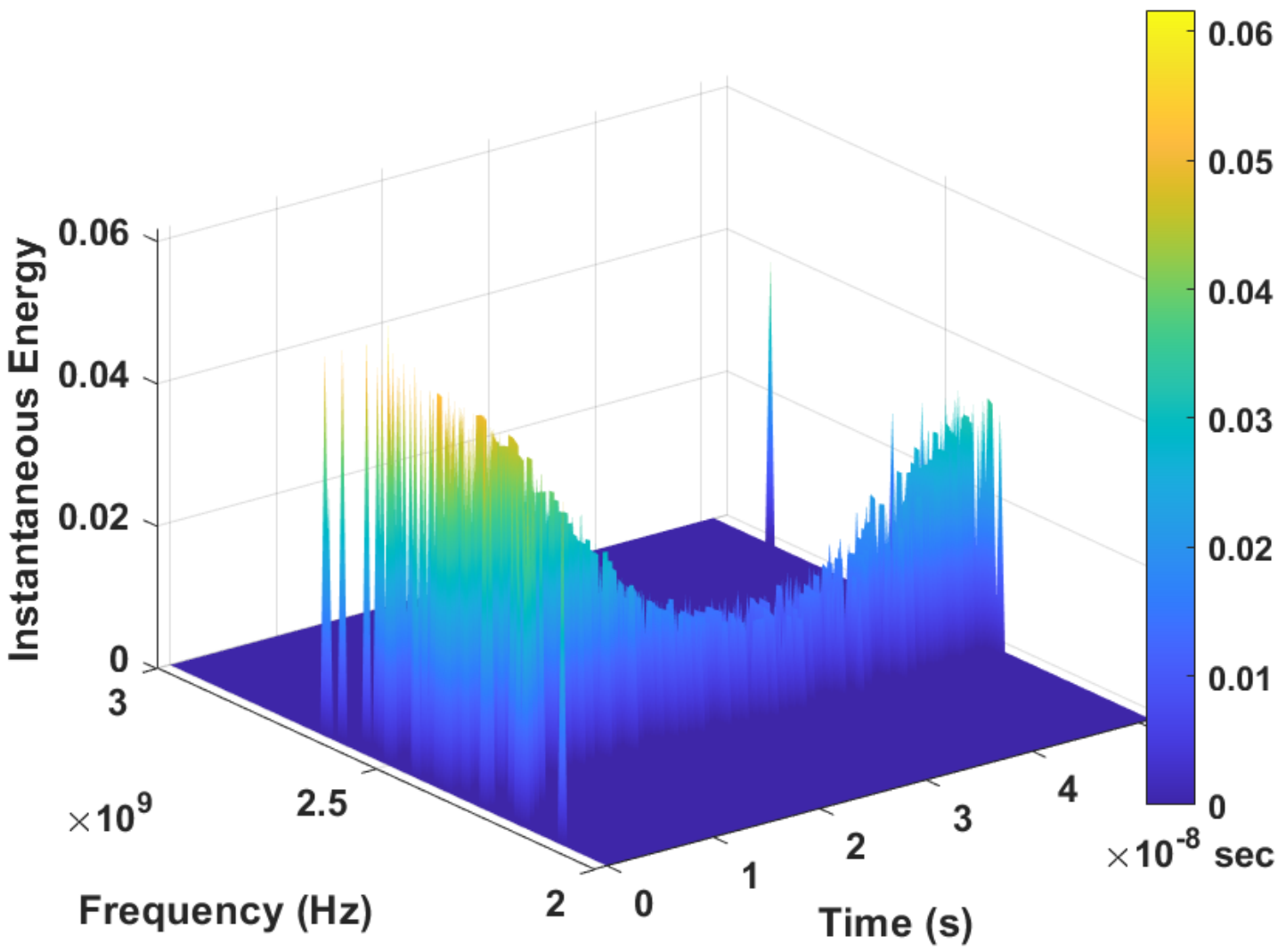}\label{}}
\end{subfigure}
\hspace{5mm}
\begin{subfigure}[]{\includegraphics[scale=0.48]{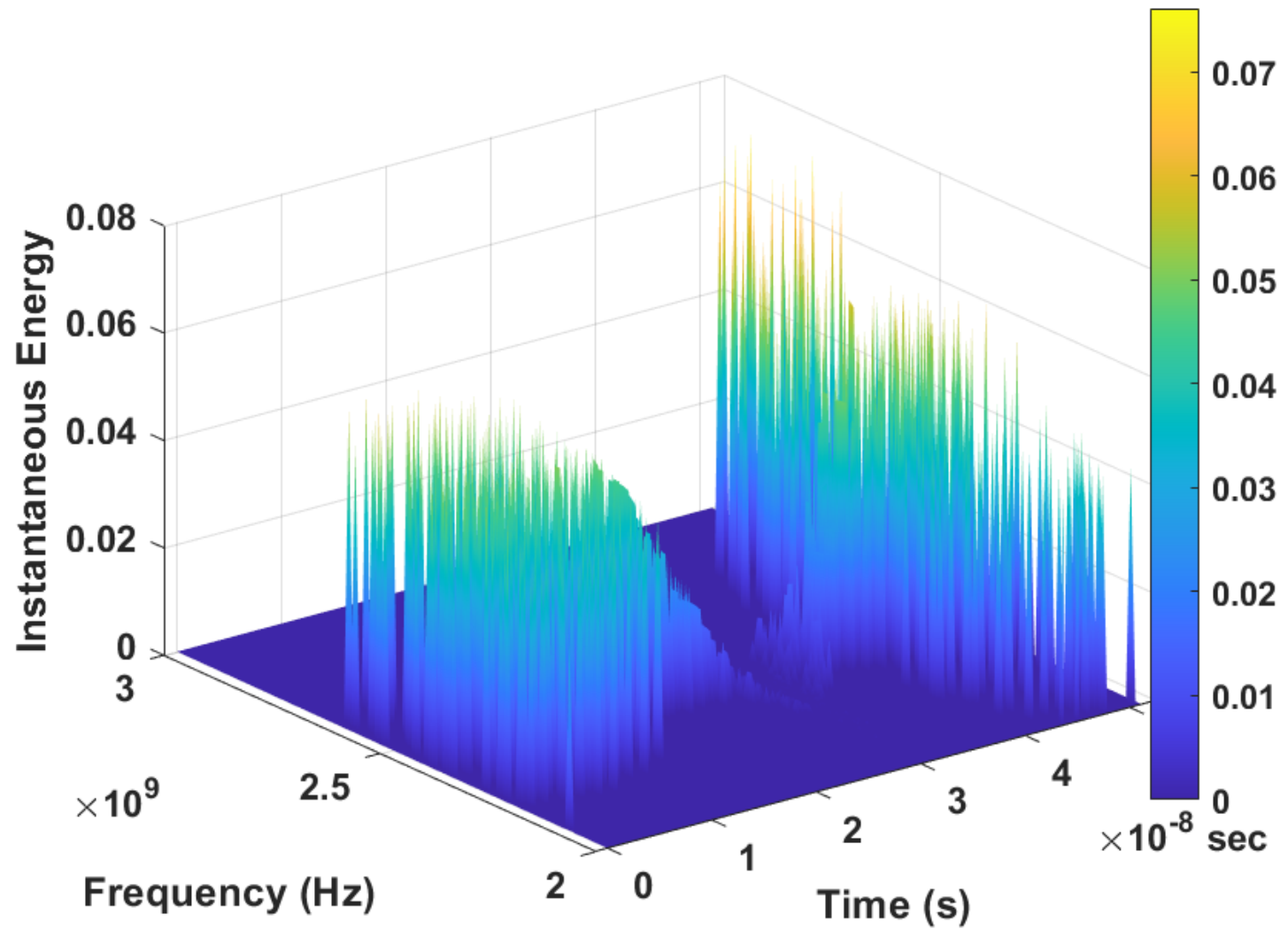}\label{}}
\end{subfigure}
 \caption{{The Hilbert spectrum showing the time-frequency-energy characteristics from the steady state of the captured RF signals: (a) DJI Inspire, (b) Beebeerun. This shows that the time-frequency-energy distribution of UAV signals varies. The variation is exploited as signatures for UAV identification.}}
  \label{uav_hht}}
  \vspace{-5mm}
\end{figure*}

EMD unveils hidden quasi-periodicity and features by decomposing a signal into a set of oscillating waves called intrinsic mode functions (IMFs) and residual based on sifting procedure \cite{huang1998empirical,kim2009emd,stallone2020new}. The signal $x(t)$ is expressed as: 
\begin {equation} \label{eq:emd}
x(t)= \sum_{i=1}^{n} {{imf(t)}_i} + r(t) ,
\end{equation}
where $n$ is the number of IMFs, $imf$ denotes intrinsic mode function and $r(t)$ represent the residual component.

HT is used for spectral analysis of each IMF to extract the frequency information with respect to time and measure the variation of the oscillating wave at different time spaces and location \cite{kim2009emd,huang1998empirical}. For each IMF, the HT returns an analytic signal $z$ as defined in (\ref{eq:ht}).
\begin {equation} \label{eq:ht}
z_i{(t)}= imf_i(t) + j H{\{imf_i(t)\}} ,
\end{equation}
where $H{\{imf_i(t)\}}$ is the HT of $imf_i(t)$. Analytic signal, $z_i(t)$, can be represented in a polar form as (\ref{eq:ht_polar}):
\begin {equation} \label{eq:ht_polar}
z_i{(t)}= a_i(t)~e^{j{\theta}_i (t) },
\end{equation}
where $a_i(t)$ is the instantaneous amplitude and ${\theta}_i (t)$ is the instantaneous phase.
Instantaneous energy is expressed as $|a_i(t)|^2$.
In this work, only the first two IMFs are used for feature extraction. For each IMF (i.e., $i={1,2}$), the instantaneous energy is statistically summarized by thirteen statistical parameters as features giving a total of twenty-six features from HHT. The thirteen statistical parameters are described in Table~\ref{Table_features}. Fig.~\ref{emd_hht} shows the EMD of signals from DJI Inspire and Beebeerun UAV with first two IMFs and the residual for each signal. The HT for the first two IMFs of signals from DJI Inspire and Beebeerun UAV are shown in Fig.~\ref{uav_hht}(a) and Fig.~\ref{uav_hht}(b), respectively. This shows the magnitude of the energy in time-frequency domain.
\begin{table}
\centering
\caption{Definition of Statistical Features used for feature extraction.}
\label{Table_features}
\begin{tabular}{|p{2.5cm}| m{2.9cm}|m{2.3cm}|}
\hline
Features & Formula & Measures\\
\hline
Mean~($\mu$)& $\frac{1}{N}\sum_{i=1}^N x_{i}$~& Central tendency \\
Harmonic mean &$(\frac{1}{N}\sum_{i=1}^N x_{i}^-1)^-1$~ & Central tendency\\
Standard deviation~($\sigma_{T}$)~& $\left[\frac{1}{N-1}\sum_{i=1}^N (x_{i}-\bar{x})^2\right]^{\frac{1}{2}}$ & Dispersion \\
Variance & $\frac{1}{N}\sum_{i=1}^N (x_{i}-\mu)^2$ & Dispersion \\
Kurtosis ($k$) & $\frac{\sum_{i=1}^N (x_{i}-\bar{x})^4}{(N-1)\sigma_{T}^4}$ & Tail/shape descriptor~\\
Root mean square ($x_{\rm{rms}}$)~& $\left[\frac{1}{N}\sum_{i=1}^N x_{i}^2\right]^\frac{1}{2}$ & Magnitude/Average power\\
Shape factor($x_{\rm{sf}}$)& $\frac{x_{\rm{rms}}}{\bar{x}}$ & Shape descriptor \\
Peak value ($x_{\rm{pv}}$)& $\rm{max(x_{i})}$ & Amplitude \\
Peak to peak ($x_{\rm{ppv}}$) & $\rm{max(x_{i})}-\rm{min(x_{i}})$ & Waveform amplitude\\
Interquantile range  &$Q_3 -Q_1$ &  Dispersion\\
Shannon entropy($H_s$) & $-\sum_{i=1}^N {x_{i}}^2 \log {x_{i}}^2$ & Uncertainty\\
Summation($S$)  & $\sum_{i=1}^N x_{i}$ & Sum of amplitude\\
\hline
\end{tabular}
% \vspace{-3mm}
\end{table}

\subsection{Wavelet Packet Transform For Time-Frequency Domain Feature Extraction}
\begin{figure}[h]
\center{\includegraphics[ clip,scale=0.5]{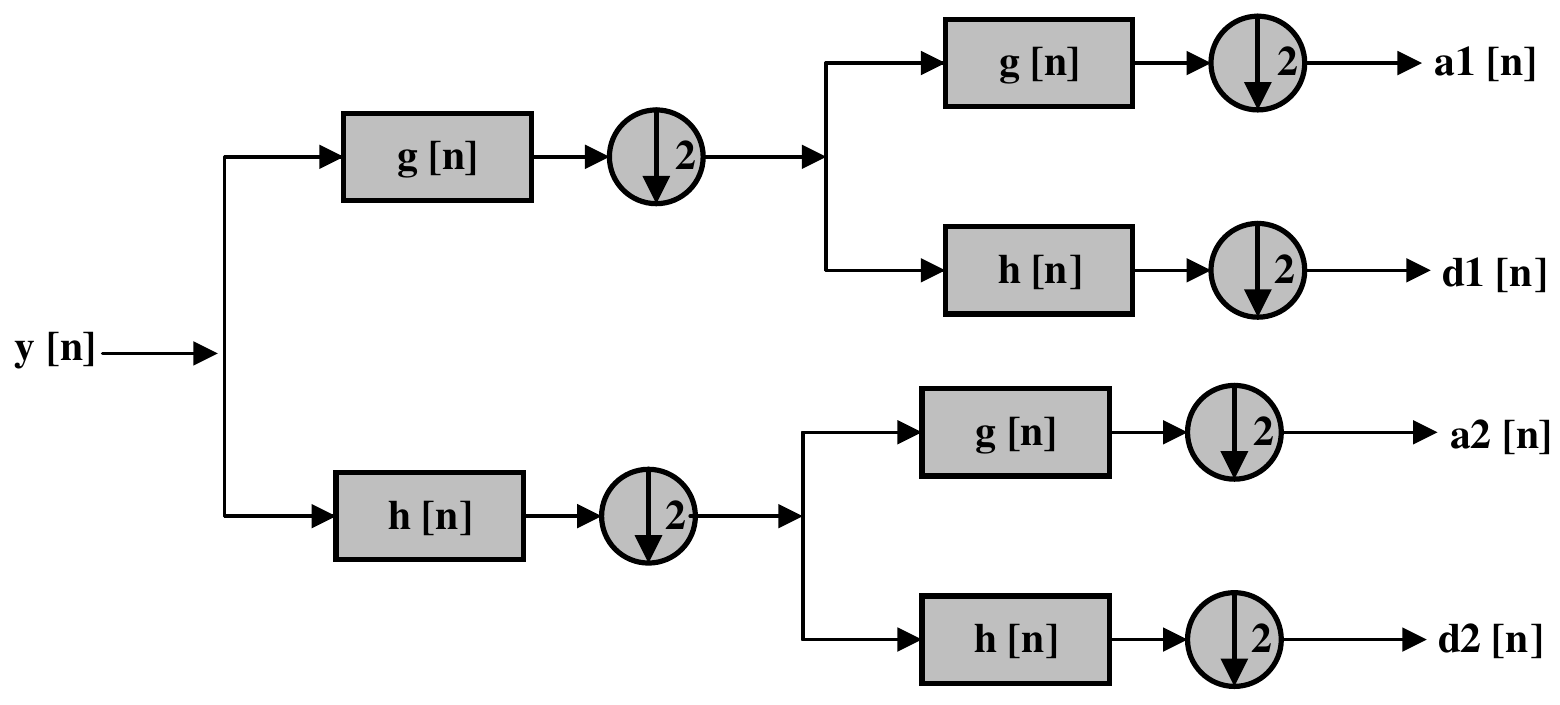}}
\caption{A two-level wavelet packet transform tree used to decompose the captured RF signals into packets for feature extraction (adopted from \cite{medaiyese2021semi}).}
\label{Fig:wpt}
\vspace{-5mm}
\end{figure}
In addition to the twenty-six features from HHT, we extracted sixteen statistical features by decomposing the signal using wavelet packet transforms (WPT). WPT is similar to the common discrete wavelet transform (DWT) where signal is sub-banded into approximation and detail coefficients \cite{wickerhauser1991lectures}. The main difference between DWT and WPT is that in WPT both the approximation and detail coefficients are decomposed further for every level of decomposition. This allows WPT to have a richer time-frequency domain analysis. However, only the approximation coefficient is decomposed further in DWT. Fig.~\ref{Fig:wpt} shows a two level WPT tree that is used for decomposing signals in this work. The two level WPT decomposes signal into four packets (i.e., $a1[n]$, $d1[n]$, $a2[n]$, and $d2[n]$). For each packet, four statistical parameters are used to describe the properties of the packet as features. Hence, sixteen features are extracted using WPT. These four statistical parameters are standard deviation, variance, peak to peak magnitude, and Shannon entropy. The definitions of these four statistical parameters are in Table~\ref{Table_features}.

\subsection{Identification Algorithms Using hierarchical learning}
We introduce the concept of hierarchical learning for the identification of UAVs. Most research works have used flat classification (i.e., binary or multi-class classification) approaches to solve the problem of UAV identification. Hierarchical classification is a learning approach where classes in a classification problem are systematically organized as a class hierarchy such that the organization forms a tree structure or direct acyclic graph (DAG) \cite{silla2011survey}. "Systematically organized" implies that classes with high similarities are categorized into meta-classes. For instance, UAVs from the same manufacturer can form a meta-class that bears the manufacturer's name. Basically, any classification problem that the classes have \textit{'IS-A'} relationship can be considered as hierarchical classification \cite {silla2011survey,wu2005learning}.
\begin{figure*}
\center{\includegraphics[clip,scale=0.56]{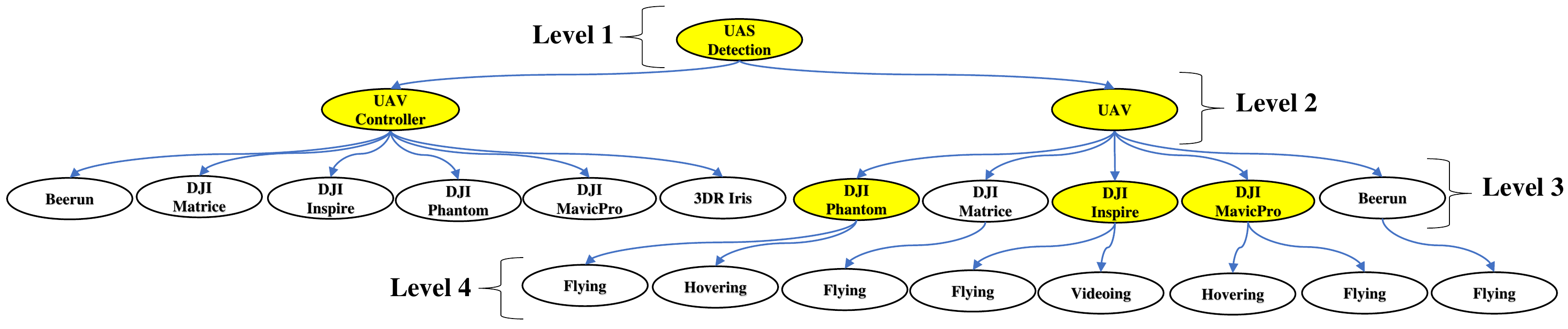}}
\caption{Tree based structure for UAS identification using hierarchical classification. The nodes in yellow are nodes with more than one child node. A classifier is employed at every node with more than one child node. For this reason, the hierarchical classifier is made up of six classifiers. }
\label{Fig:hierarchy_structure}
 \vspace{-5mm}
\end{figure*}
According to  \cite{silla2011survey, freitas2007tutorial,sun2001hierarchical}, three criteria  differentiate one hierarchical classification method from another. The first criterion is the hierarchy structure which can be a tree-based or DAG \cite{silla2011survey}. The disparity between a tree-based and DAG is that a node could have more than one parent node in DAG. The second criterion is the levels at which classification is performed. If the hierarchical classifier is designed to only classify the leaf node of the hierarchy then it is referred to as mandatory leaf-node prediction (MLNP) \cite{freitas2007tutorial}. Conversely, if the classifier is implemented to predict any node at any level in the hierarchy then such method is called non-mandatory leaf-node prediction (NMLNP). The third criterion is the mannerism of exploring the hierarchical structure \cite{silla2011survey}. These can be: (i) a top-down approach where local classifiers are utilized at every node or  parent node or level in the hierarchy; (ii) global approach where a single classifier is designed to handle all the class hierarchy; (iii) flat approach which disregards the hierarchy relationship among classes.

% \subsection{Local Hierarchical Classification (LHC)}
Local hierarchical classification (LHC) is a top-down approach that can utilize local classifier at a node, parent node or level in a hierarchy by taking advantage of local information to make a better classification \cite{silla2011survey}. The setback of LHC is that error (misclassification) at a level will be propagated to the subsequent lower levels in the hierarchy. There are three ways of implementing and they include:
\begin{enumerate}
    \item a local classifier per node (LCN): A binary classifier is used for every node in the hierarchy. 
    \item a local classifier per parent node (LCPN): A classifier is used for every parent node in the hierarchy to classify the children nodes. A multi-class classifier or binary classifier can be adopted depending on the number of children nodes or use case. 
    \item a local classifier per level (LCL): A single classifier for every level in the hierarchy to classify nodes in the given level.
\end{enumerate}

According to \cite{silla2011survey}, the algorithm for hierarchical classification is described by four-tuple as shown:

\begin {equation} \label{eq:4}
HC= \langle \Delta, \Xi, \Omega, \Theta \rangle, 
\end{equation}
 
where 
\begin{itemize}
    \item $\Delta$ defines the ability of the algorithm to predict single or multiple paths in the hierarchy.  This implies that this variable can only take two values which are single path prediction (SPP) and multiple path prediction (MPP).
    \item $\Xi$ defines the level or depth at which classification is performed. As previously discussed, this can be either MLNP or NMLNP.
    \item $\Omega$ is the hierarchy structure which can be a tree-based (T) or DAG.
    \item $\Theta$ defines the mannerism of exploring the hierarchical structure. This can be local or global. If local approach is adopted then the variable can be LCN, LCL or LCPN.
\end{itemize}

A new variable $\Lambda$ is introduced in the description of the HC algorithm. This makes the HC algorithm to be described by five-tuple in (\ref{eq:5}).

\begin {equation} \label{eq:5}
HC= \langle \Delta, \Xi, \Omega, \Theta, \Lambda \rangle, 
\end{equation}

where $\Lambda$ is the ML classification algorithm. This can be support vector machine (SVM), deep neural network (DNN), logistic regression,  XGBoost, decision tree, ensemble and so on. Hence, our proposed HC algorithm in this work is defined as in (\ref{eq:6}). Introducing $\Lambda$ gives a complete information and definition of a HC algorithm.

\begin {equation} \label{eq:6}
HC= \langle MPP, NMLNP, T, LCPN, XGBoost \rangle, 
\end{equation}

LCPN reduces inconsistent predictions and gives tolerance for class relationships \cite{silla2011survey}. Tree-based structures simplify the class relationship. MPP helps in understanding the prediction flow and NMLNP enables the adaptability of the algorithm for various use case. XGBoost is the ML classifier which is a scalable tree boosting algorithm that sequentially combines several predictors (i.e., decision trees) \cite{chen2016xgboost}. Each predictor improves on the error of its predecessor. Detailed information about XGBoost can be found in \cite{chen2016xgboost}.

Fig.~\ref{Fig:hierarchy_structure} shows the tree hierarchy of the UAS classification problem. The first level (i.e., level 1) has the root node. The root node has two child nodes that categorize a UAS into a UAV controller or a UAV. These child nodes forms the second level (i.e., level 2) of the hierarchy. The second level categorizes the model type for the UAVs or UAV controllers. The classification of the model type forms the third level (i.e., level 3). The third level categorizes the flight mode of UAVs. The flight mode of UAVs form the fourth level of the hierarchy. At every parent node with more than one child node, we trained an XGBoost classifier to class the child nodes. The classifiers at all parent nodes (i.e., nodes in yellow) are cascaded as shown in Fig.~\ref{Fig:hierarchy_structure}.

\section{Experimental Results and Discussions} \label{seven}
We evaluated the proposed framework by considering the five main directions. These directions are:
\begin{itemize}
\item The performance of SDAE based on the reconstruction error
\item The performance of LOF based on: i) the selection of the number of nearest neighbors for determining local density; ii) the type of distance measure; iii) classification metrics; iv) amount of training data required; v) inference time; vi) impact of channel effect.
\item The performance of each composing model of the hierarchical classifier.
\item The performance of the LHC based on hierarchical classification metrics (i.e., hierarchical precision, hierarchical recall, and hierarchical F1-score).
\item The effect of feature set from the WPT on the LHC's performance.
\end{itemize}

\subsection{Evaluation of SDAE}
\begin{figure}[h!]
\center{\includegraphics[ clip,scale=0.5]{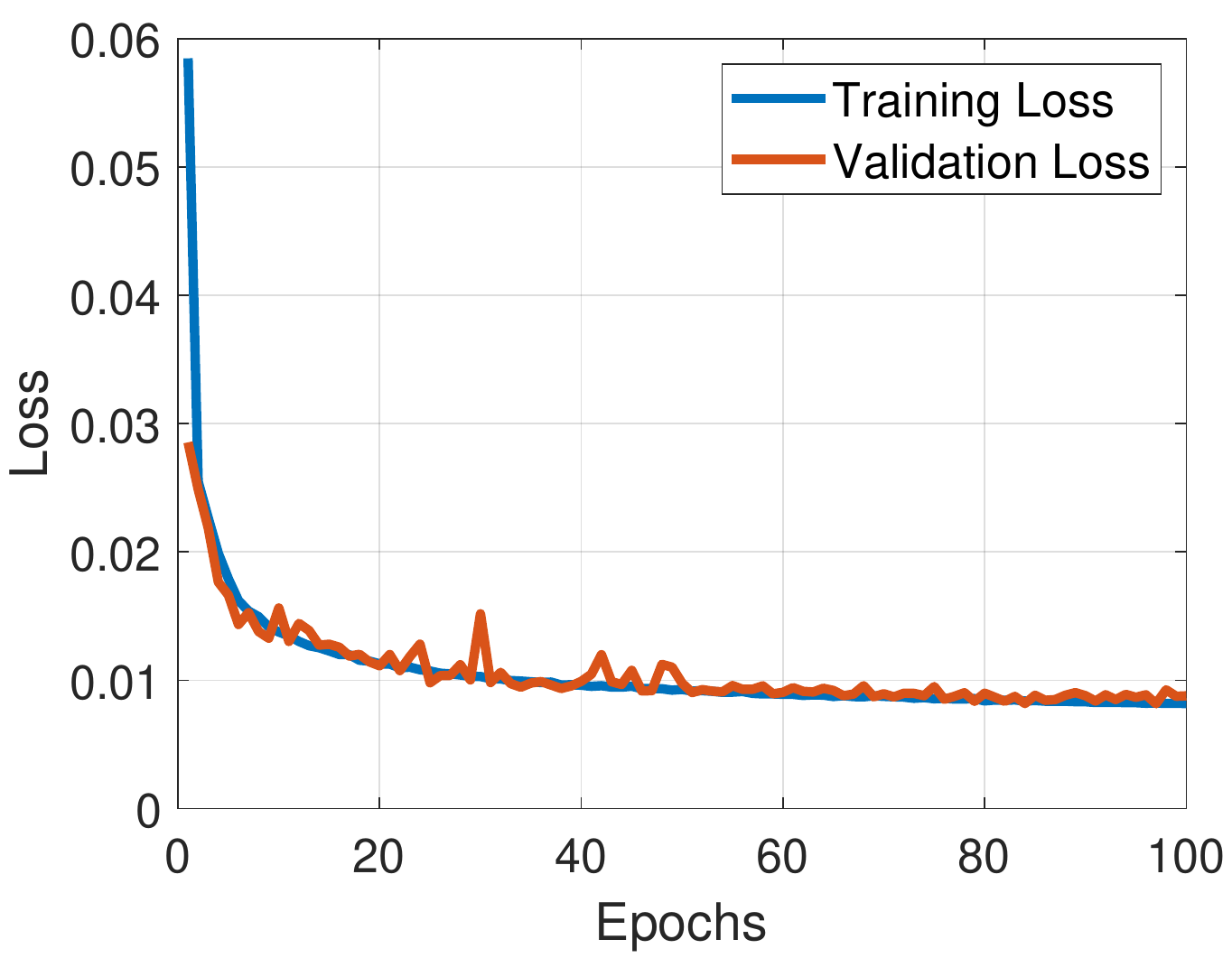}}
\caption{The learning curve of SDAE. This shows learning performance of the SDAE at every epoch. The loss function of the training set and validation set drop steadily per epoch.}
\label{Fig:sdae_learning_curve}
\vspace{-3mm}
\end{figure}
The SDAE is trained using certain hyperparameters. Mean absolute error (MAE) is used to estimate the reconstruction error or as the loss function. The Adam optimization is used as the optimizer. The learning rate, epoch and batch size are 0.001, 100 and 128 respectively. The learning curve of the SDAE is shown in Fig.~\ref{Fig:sdae_learning_curve} where the reconstruction error for both training and validation sets decrease steadily per epoch. The test set for evaluating SDAE comprises signals from Bluetooth and WiFi devices the same way as the training set is consisting of signals from Bluetooth and WiFi devices. However, the test set is not corrupted with AWGN (i.e., the test set's signals are at 30 dB SNR). When SDAE was evaluated using the test set, the total reconstruction error (MAE) is 0.00793. This implies that the SDAE was able to reconstruct the all the test set with approximate error equal to zero. 

\subsection{Evaluation of LOF}
The LOF model can be evaluated like a binary classifier because it classifies a signal either as an inlier or outlier. So, classification metrics such as accuracy, precision, recall, and F1-score can be used to evaluate the performance of the classifier. The definition of these metrics is given below:
\begin {equation} \label{eq:7}
\text{Accuracy}=\frac{ T_{\rm{P}}+T_{\rm{N}}} {T_{\rm{P}}+T_{\rm{N}}+F_{\rm{P}}+F_{\rm{N}}},
\end{equation}

\begin {equation} \label{eq:8}
\text{Precision}= \frac{T_{\rm{P}}}{T_{\rm{P}}+F_{\rm{P}}},
\end{equation}

\begin {equation} \label{eq:9}
\text{Recall}= \frac{T_P}{T_P+F_N} ,
\end{equation}

\begin {equation} \label{eq:10}
{F_{\rm{1}}}~\text{score}=2\left( \frac{Precision\times Recall}{Precision + Recall}\right),
\end{equation}

\noindent where $T_{\rm {P}}$, $T_{\rm {N}}$, $F_{\rm {P}}$ and $F_{\rm {N}}$ represent true positive, true negative, false positive and false negative, respectively.

After training the SDAE, the decoder is detached from the network and left with the encoder. This is because we only use the encoder of the SDAE to reduce the signal into a latent representation. The training set is passed into the encoder of the SDAE to compress the signal. The compressed training set is used to train the LOF algorithm. The evaluation set for the LOF model consists of 100,000 recognized signals (i.e., Bluetooth  and WiFi) and 100,000 UAS signals. The evaluation set is split into test and validation sets in the ratio of 80\% to 20\%, respectively. 

The two major hyperparameters of LOF are the distance measure and number of nearest neighbors.
The distance measure is used for measuring similarity in data. There are different distance metrics, and these include Manhattan (also known as $L1$ or city block), Euclidean (also known as $L2$), Cosine distance, Minkowski distance, and so on. For training purpose, 
Manhattan distance is used for the distance measure. However, we assess the impact of the distance measure on the model's performance using Euclidean and Cosine distance.

\begin{figure}
\center{\includegraphics[ clip,scale=0.5]{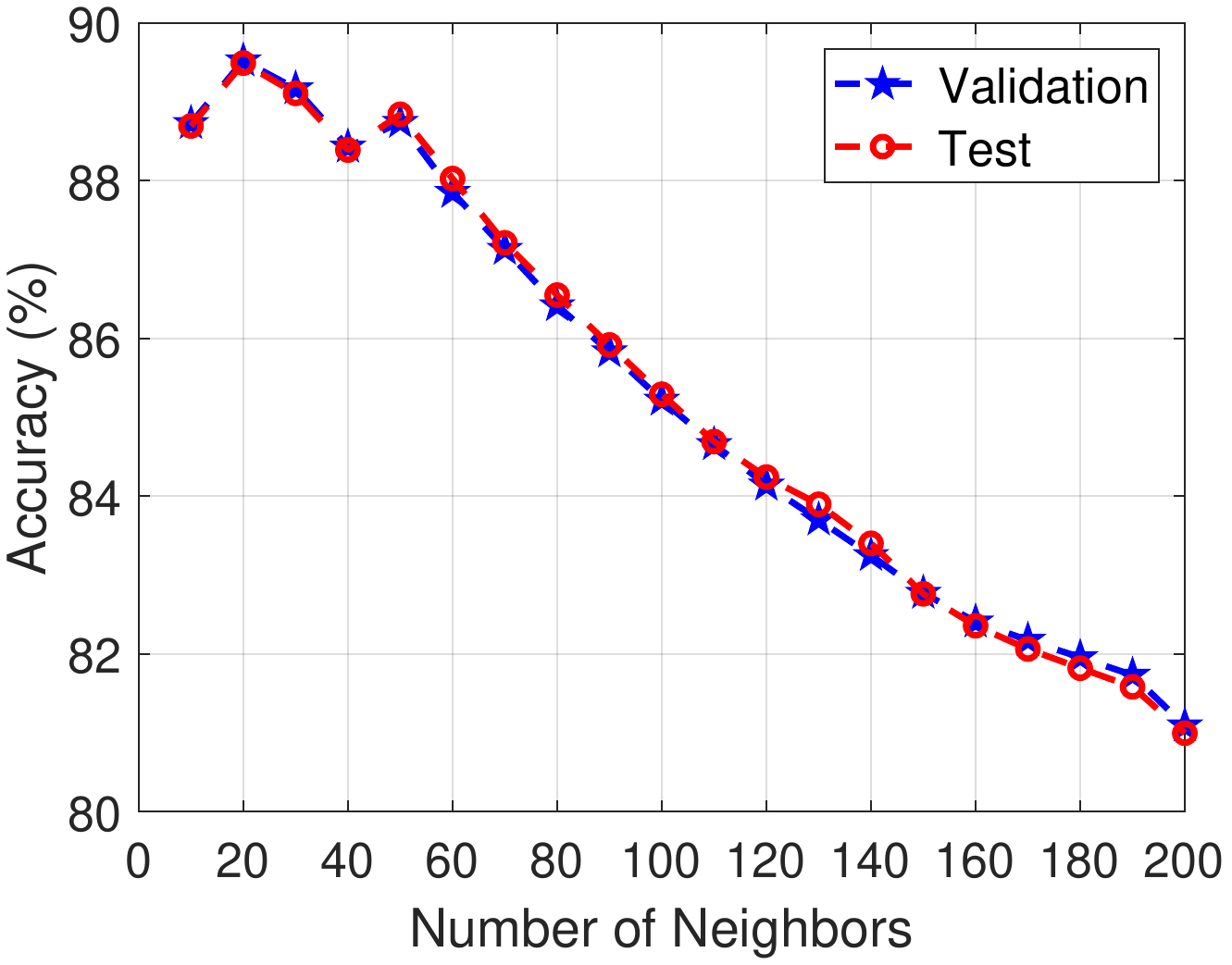}}
\caption{ Average accuracy of LOF model based on the validation and test set when increasing the number of nearest neighbors for local density estimation at 30 dB SNR. }
\label{Fig:sdae_lof_test_val}
\vspace{-3mm}
\end{figure}

Determining the number of nearest neighbors to estimate local density of a datapoint is not a trivial task when using LOF because it affects the performance of the model. So, the validation set is used in selecting the appropriate number of nearest neighbors. The number of nearest neighbors is initially set at 10 and iteratively increases by a step of 10 until 200. At every iteration the validation set is used to assess the performance of the model based on classification accuracy. Fig.~\ref{Fig:sdae_lof_test_val} shows the performance of the LOF model as the number of nearest neighbors increases. When the number of nearest neighbors $n$ is 20, the highest classification accuracy of 89.49\% was achieved using the validation set. As we increase the number of nearest neighbors from 20, the model's accuracy continues to drop. For this reason, the LOF model uses 20 as the number of neighbors to estimate the local density of a data point.

\begin{figure}[h]
\center{\includegraphics[ clip,scale=0.45]{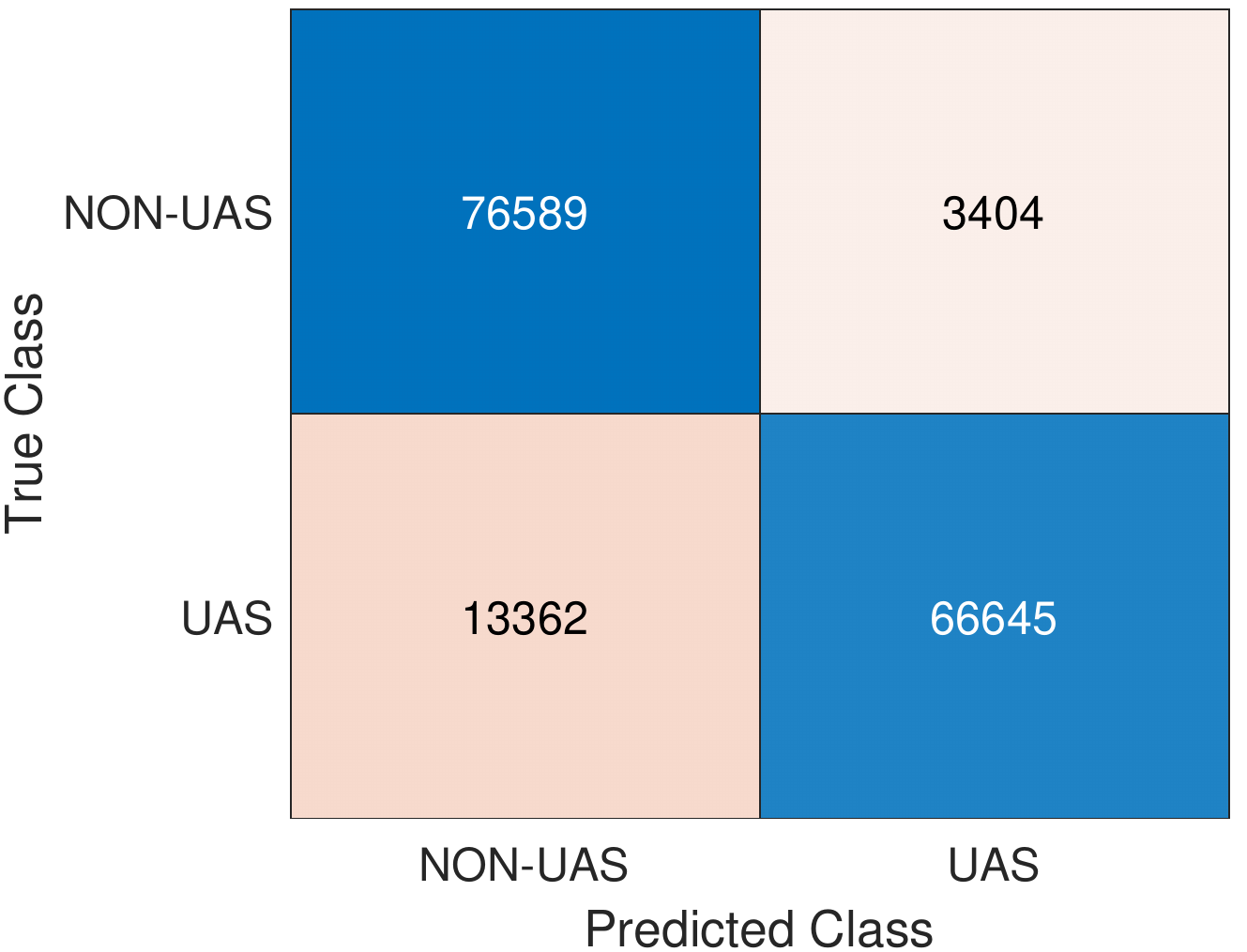}}
\caption{ Confusion matrix of the LOF model ($k=20$) when evaluating the model with the test set at an SNR of about 30~dB. }
\label{Fig:sdae_lof_confusion_matrix}
\vspace{-1mm}
\end{figure}

\begin{figure*}
\center{
\begin{subfigure}[]{\includegraphics[scale=0.48]{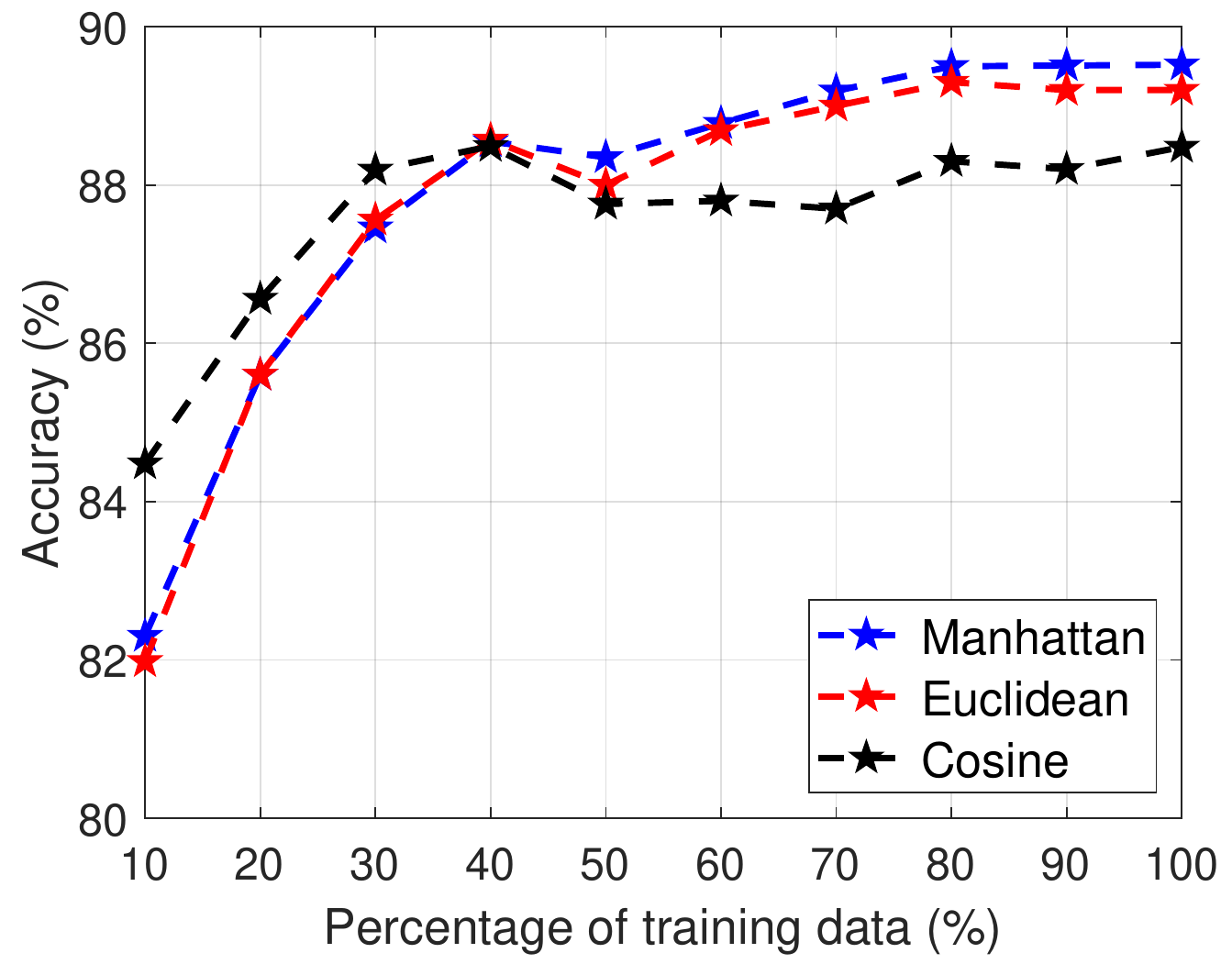}\label{}}
\end{subfigure}
\hspace{5mm}
\begin{subfigure}[]{\includegraphics[scale=0.48]{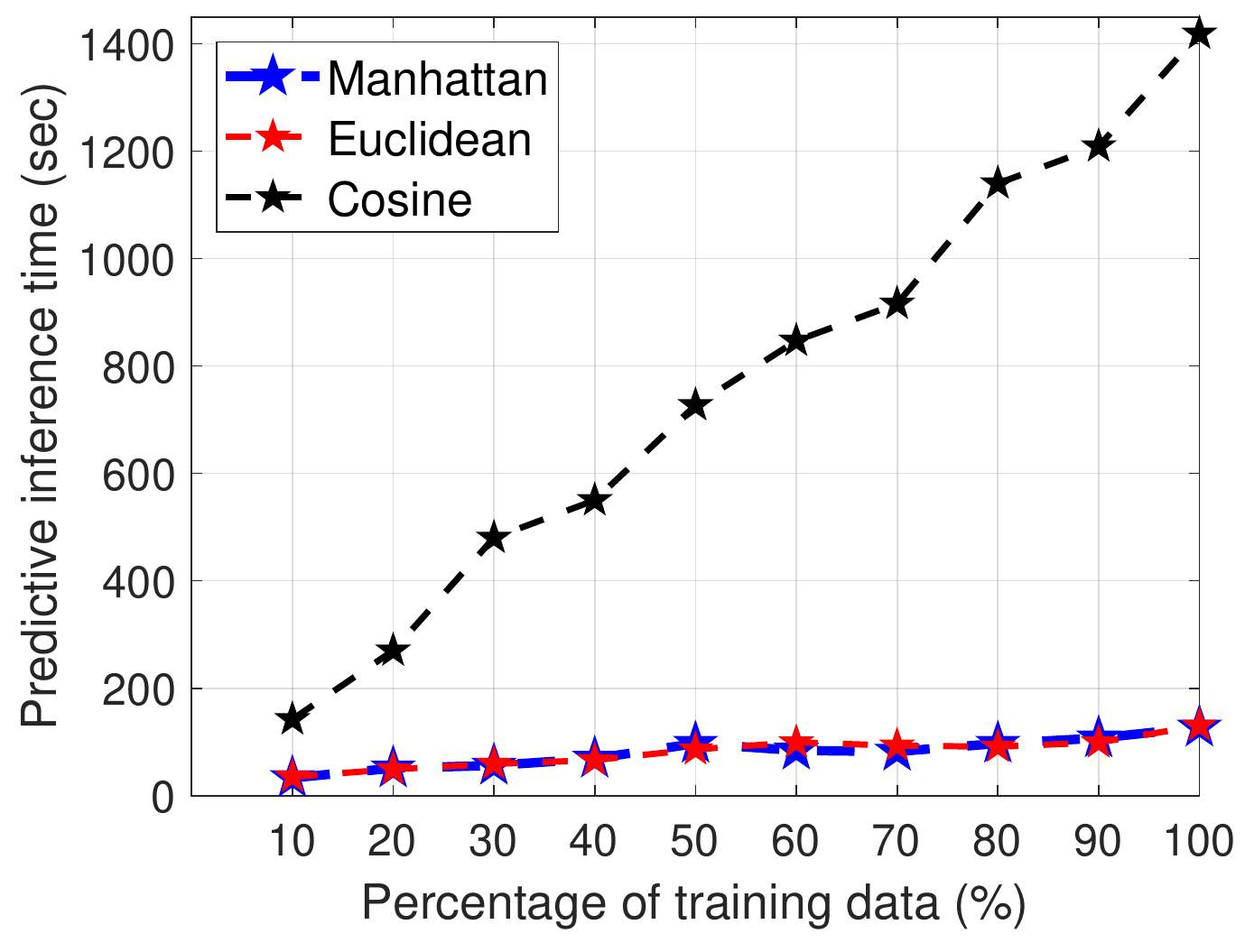}\label{}}
\end{subfigure}
 \caption{ (a) The relationship between training set size and accuracy for the LOF model, (b) The relationship between training set size and inference time for the LOF model. Using either Manhattan or Euclidean distance have no significant differences on the LOF model. The Manhattan and Euclidean distance outperforms the Cosine distance. }
  \label{train_size_variation}}
  \vspace{-5mm}
\end{figure*}

The performance of the LOF classifier is assessed using the test set. The confusion matrix of the classifier using the test set is shown in Fig.~\ref{Fig:sdae_lof_confusion_matrix}. The confusion matrix's rows refer to the true class and the columns show the predicted class. The LOF classifier is able to classify 95.7\% of the non-UAS signals and 83.3\% UAS signals as the true class.

Furthermore, the total number of training examples is 234,500 signals. So, the impact of training set size on accuracy and predictive inference time is assessed by varying the percentage of training data and the distance metric (i.e., using Euclidean and Cosine distance as a substitute for Manhattan distance). When an unknown signal is sent into the model, the predictive inference time is the amount of time it takes the LOF classifier to infer the signal type in the test set.

Fig.~\ref{train_size_variation}(a) shows the relationship of classification accuracy, distance metric, and size of the training set. As the training set increases, the accuracy of the LOF model increases when using Manhattan and Euclidean distance. Both Manhattan and Euclidean distance have a comparable performance using 10\% to 40\% of the training data. As we increase the training data from 40\% upward, the Manhattan distance slightly yields a higher accuracy. Similarly, the Cosine distance gives an accuracy higher than both Manhattan and Euclidean distance when the 30\% or less training data is used. However, both Manhattan and Euclidean distance performs better than Cosine distance when the training data increases. 

Fig.~\ref{train_size_variation}(b) depicts the relationship of predictive inference time, distance metric, and size of the training set for the LOF classifier. As the training set increases, the predictive inference time increases. This is because the LOF classifier is a non-parametric model. It calculates the local density of each data point and assigns values that represent the degree of being an outlier. Both Manhattan and Euclidean distance also have a comparable predictive inference time. On the other hand, the Cosine distance takes a higher time than both Manhattan and Euclidean distances. Using all the training data, the average predictive inference time for Manhattan and Euclidean distance is below 150 seconds while Cosine distance takes over 1400 seconds. The use of Cosine distance is not time efficient compare to the other two distance measure.

In addition, the effect of channel noise on the LOF model is examined by varying the SNR of the signals. AWGN is added to the signals to vary the SNR. This assessment is done by examining the impact of the number of nearest neighbors and distance measure (i.e., using Euclidean and Cosine distance instead of Manhattan distance) with respect to varying SNR.

\begin{figure}
\center{\includegraphics[ clip,scale=0.52]{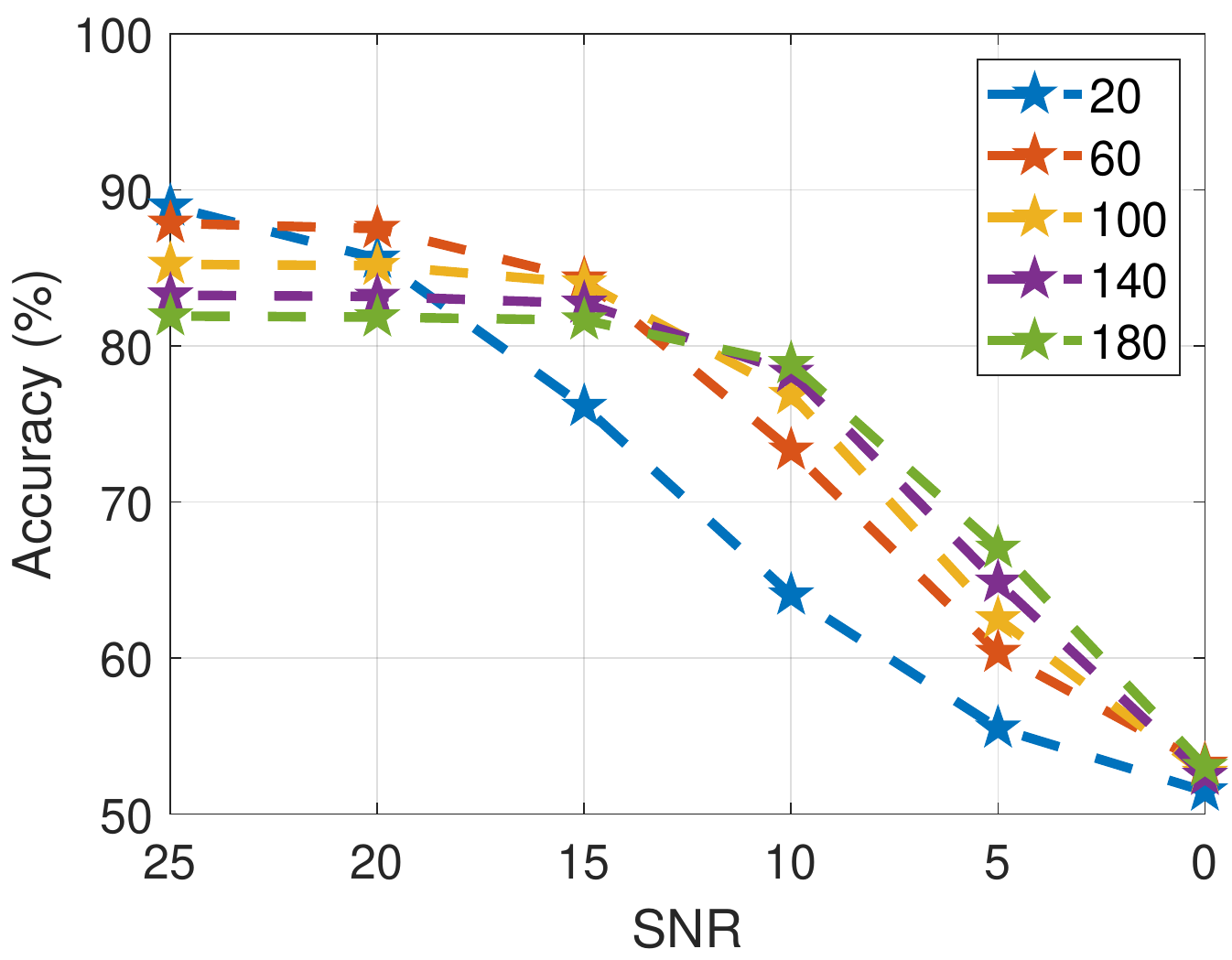}}
\caption{ The classification accuracy of LOF model (distance measure is Manhattan) under varying SNR and the number of nearest neighbors. The number of nearest neighbors is increased at the step of 40 starting from 20 to 180. }
\label{Fig:sdae_lof_snr_accuracy}
\vspace{-5mm}
\end{figure}

\begin{figure}
\center{\includegraphics[ clip,scale=0.5]{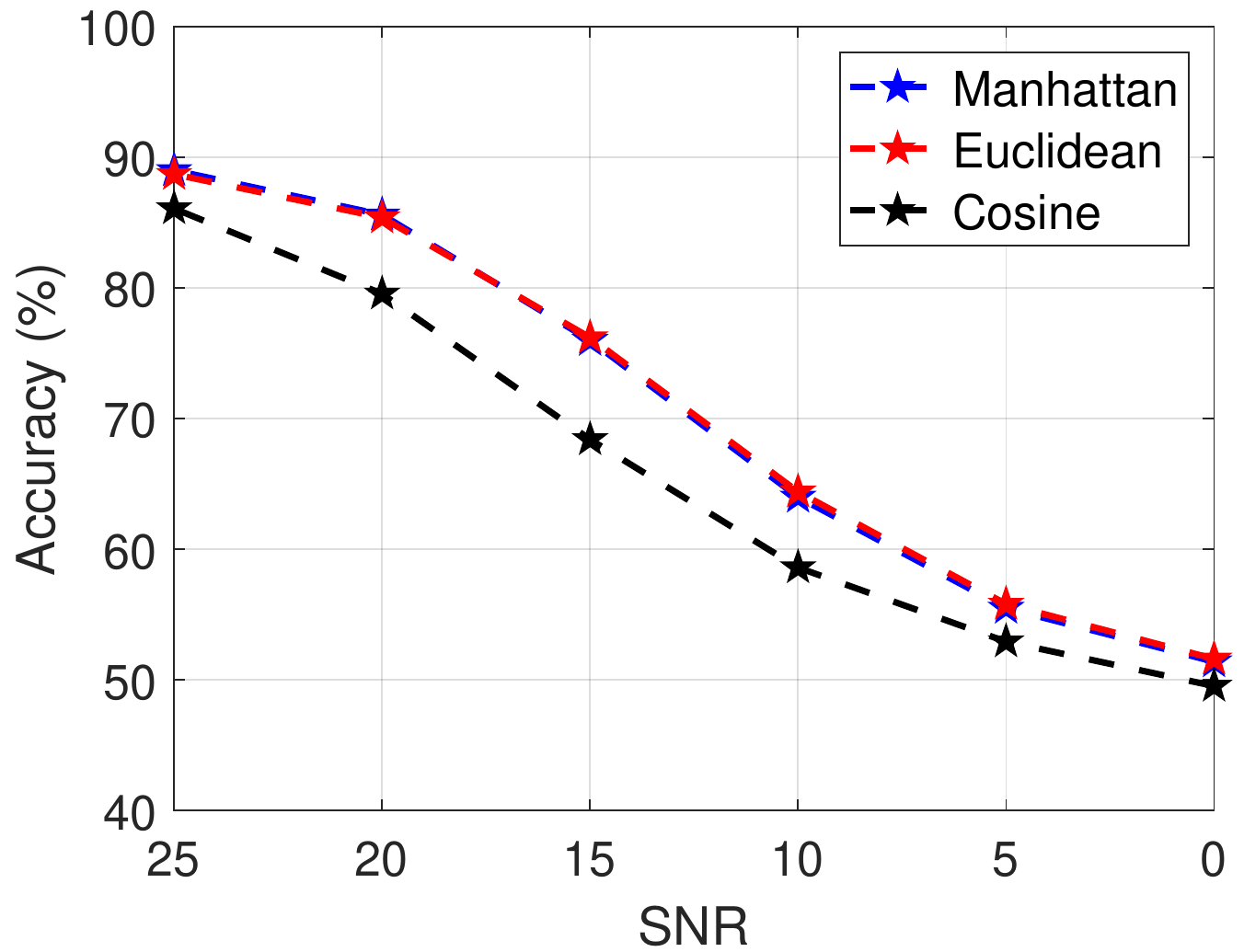}}
\caption{Performance of the LOF model (the number of nearest neighbors is 20) under varying SNR and distance measure. The Manhattan and Euclidean distance have similar characteristics when varying the SNR of the signal. Both Manhattan and Euclidean distance outperforms the Cosine distance.  }
\label{Fig:sdae_lof_snr_distance}
\vspace{-5mm}
\end{figure}

Fig.~\ref{Fig:sdae_lof_snr_accuracy} shows the behavior of the model under varying SNR and the number of nearest neighbors. The accuracy of the model drops gradually with a decrease in the signal's SNR. At 25~dB, using 20 as the number of nearest neighbors, the model's classification accuracy is about 89\%. This performs better than increasing the value of the number of nearest neighbors. However, from 15~dB to 0~dB, the model's performance improves with an increase in the number of nearest neighbors. For instance, at 10 dB, when the number of nearest neighbors is 20, 60, 100, 140, and 160, the model's accuracy is 64\%, 73\%, 77\%, 78\%, and 79\%, respectively. Fig.~\ref{Fig:sdae_lof_snr_distance} depicts the performance of the model when varying the SNR using the three distance measures. The Manhattan and Euclidean distance have similar characteristics, and they outperform the use of Cosine distance at different levels of SNR.

\subsection{Hierarchical Classifier  Performance based on classification metrics}
\begin{table*}
\setlength{\tabcolsep}{1.3pt}
\centering
\caption{Performance measure of each local classifier in the hierarchy.}
% \tablefootnote{vvvv}
\label{Table_xgb_model}
% \begin{tabular}{p{2cm}| m{2cm}|m{2cm}|m{2cm}|m{2cm}|m{2cm}}
\begin{tabular}{|c|c|c|c|c|c|c|c|c|c|c|}
\hline
 Level & Model & \multicolumn{4}{c|}{HHT}  &  \multicolumn{4}{c|}{HHT-WPT}& Accuracy   \\    \cline{3-10}
& & Accuracy (\%) & Precision (\%) &Recall (\%) & $F_1$-score (\%) & Accuracy (\%)& Precision (\%)& Recall (\%)& $F_1$-score (\%)&difference(\%)\\ 
 \hline
1 & UAS classifier & \textbf{87.10} & 86.20 & 86.37 & 86.28 &  \textbf{91.00} & 90.23 & 90.60 & 90.42 & \textbf{3.90} \\
\hline
2& UAV controller classifier & \textbf{71.19}&	72.53&	73.16&	72.72  &  \textbf{73.19} & 74.01 &74.30 &	74.06& \textbf{2.00}\\
\hline
2 & UAV classifier& \textbf{81.45}&	81.19&	81.46&	81.27  &  \textbf{82.49} &	82.18 &	82.50 &	82.25 &	 \textbf{1.04}\\
\hline
3 & DJI Phantom classifier  & \textbf{87.37}&	88.62 &	86.00&	87.30 &  \textbf{88.58} &	89.71 &	87.40 &	88.54 & \textbf{1.21}	 \\
\hline
3 & DJI Inspire classifier& \textbf{94.56}&	95.70&	93.28&	94.50  &  \textbf{95.28} & 95.98 &	94.45 &	95.21 & \textbf{0.72}	 \\
\hline
3 & DJI MavicPro classifier & \textbf{81.40} & 86.77 & 74.52&	80.18  &  \textbf{82.39}  &	88.12	 & 75.17 &	81.13 &  \textbf{0.94} \\
\hline
\end{tabular}
 \vspace{-3mm}
\end{table*}

\begin{table*}[t]
\setlength{\tabcolsep}{1.5pt}
\centering
\caption{Performance measure of the local hierarchical classifier at different levels in the hierarchy.}
\label{Table_hierarchical_model}
\begin{tabular}{|c|c|c|c|c|c|c|c|}
\hline
 Level & \multicolumn{3}{c|}{HHT}  &  \multicolumn{3}{c|}{HHT-WPT} & Precision   \\    
 \cline{2-7}
&  Precision (\%) &Recall (\%) & $F_1$-score (\%) & Precision (\%)& Recall (\%)& $F_1$-score (\%)&difference(\%)\\ 
 \hline
1 &  \textbf{86.20}	&86.37&	86.28 &  \textbf{90.23}  & 90.60 & 90.42& \textbf{4.03} \\
\hline
2 & \textbf{64.88}	&64.88&	64.88 &  \textbf{67.86} & 67.86  & 67.86& \textbf{2.98}\\
\hline
3 &  \textbf{54.38}	&54.36&	54.37&  \textbf{56.86} &	56.90 &	56.88 &	\textbf{2.48}\\
\hline
\end{tabular}
 \vspace{-3mm}
\end{table*}

There are no standard ways of evaluating hierarchical classifier \cite{silla2011survey}. However, for a local hierarchical classification, testing is systemically carried out such that the first level prediction is used to narrow down the second level prediction and so on. This is recursively performed until it reaches the leaf node or a specific terminal node. 

The LHC is made up of six XGBoost models as shown in Fig.~\ref{Fig:hierarchy_structure}. The performance of each model is shown in Table~\ref{Table_xgb_model} using the classification metrics defined in (\ref{eq:7}) to (\ref{eq:10}). First, the classifiers are trained using the 26 features for HHT. The accuracy of the UAS classifier, UAV controller classifier, UAV classifier, DJI Phantom classifier DJI inspire classifier, and DJI MavicPro classifier is 87.10\%, 71.19\%, 81.45\%, 87.37\%, 94.46\%, and 81.40\%, respectively. To improve the accuracy of the classifiers, 16 features from WPT are used to augment the 26 features from HHT. By using the feature set from HHT-WPT, the performance of the classifiers increases. The accuracy of the UAS classifier, UAV controller classifier, UAV classifier, DJI Phantom classifier DJI inspire classifier, and DJI MavicPro classifier increased to 91\%, 73.19\%, 82.49\%, 88.58\%, 95.28\%, and 82.39\%, respectively. The effect of the WPT feature set is more pronounced in the classifier (i.e., UAS classifier) at the first level of the hierarchy. The accuracy of the UAS classifier increased by 3.9\%. Similarly, an accuracy increase of 2\%, 1.04\%, 1.21\%, 0.72\%, and 0.94\%  was achieved for the UAV controller classifier, UAV classifier, DJI Phantom classifier DJI inspire classifier, and DJI MavicPro classifier, respectively. This shows that the feature set from WPT improves the accuracy of the individual classifier in the hierarchy.

Flat classification metrics are mostly used for evaluating the performance of a hierarchical classifier. However, errors at different levels should not have the same weight or penalty \cite{silla2011survey}. In \cite{kiritchenko2006learning}, three hierarchical classification metrics that penalize the hierarchy classifier's errors at different levels are proposed. These are hierarchical precision ($hP$), hierarchical recall ($hR)$, and hierarchical $F_1$-score ($hF_1$). The definition of these metrics are:

\begin {equation} \label{eq:11}
hP= \frac {\sum_{i}|{P\;'_i \cap T\;'_i}|}  {\sum_{i}|P\;'_i|} ,
\end{equation}

\begin {equation} \label{eq:12}
hR= \frac {\sum_{i}|{P\;'_i \cap T\;'_i}|}  {\sum_{i}|T\;'_i|} ,
\end{equation}

\begin {equation} \label{eq:13}
hF_1= \frac {({\beta)^2} +1) \cdot  hP  \cdot hR}  {{\beta^2} \cdot hP+hR} , \beta \in [0, +\infty]
\end{equation}

where $P\;'_i$ is a set of predicted class(es) for test sample $i$ and its ancestors,  $T\;'_i$ is a set of true class(es) for test sample $i$ and its ancestors, $\beta$ is equal to 1 provided precision and recall are equal weight.

% \begin{table}[t!]
% \setlength{\tabcolsep}{3pt}
% \centering

% \caption{.}
% % \tablefootnote{vvvv}
% \label{Table_accuracy}
% % \begin{tabular}{p{2cm}| m{2cm}|m{2cm}|m{2cm}|m{2cm}|m{2cm}}
% \begin{tabular}{|c|c|c|c|c|c|c|c|}

% \hline
%  Model & \multicolumn{7}{c|}{number of features}\\
% \cline{2-8}  
%  &  6 & 12 & 18  & 24 & 30 & 36 & 42 \\

% \hline
% A  &
%  &
%  &
%  &
%  &
%  &
%  &

%  \\ 	

% \hline
% B  &
%  &
%  &
%  &
%  &
%  &
%  &

%  \\ 
%  \hline
% C  &
%  &
%  &
%  &
%  &
%  &
%  &

%  \\ 
%  \hline
% D  &
%  &
%  &
%  &
%  &
%  &
%  &

%  \\ 
%  \hline
% E  &
%  &
%  &
%  &
%  &
%  &
%  &

%  \\ 
%  \hline
% F  &
%  &
%  &
%  &
%  &
%  &
%  &

%  \\ 
%   \hline
% Average accuracy  &
%  &
%  &
%  &
%  &
%  &
%  &

%  \\ 

%  \hline
% \end{tabular}
%  %\vspace{-3mm}
% \end{table}

% \begin{table}
% \setlength{\tabcolsep}{3pt}
% \centering
% \caption{Performance measure of each local classifier in the hierarchy.}
% \label{Table_xgb_model}
% \begin{tabular}{|c|c|c|c|c|}

% \hline
%  Level & Accuracy (\%) & Precision (\%) & Recall (\%) & $F_1$-score (\%)\\
 
% \hline
% UAS classifier & 91.00 & 90.23 &	90.60 &	90.42

%  \\ 	

% \hline
% UAV controller classifier  & 73.19 & 74.01 &	74.30 &	74.06

%  \\ 
%  \hline
% UAV classifier  & 82.49 &	82.18 &	82.50 &	82.25

%  \\ 
%  \hline
% DJI Phantom classifier  & 88.58 &	89.71 &	87.40 &	88.54

%  \\ 
%   \hline
%  DJI Inspire classifier  & 95.28 &	95.98 &	94.45 &	95.21

%  \\ 
%   \hline
% DJI MavicPro classifier  & 82.39  &	88.12	 & 75.17 &	81.13

%  \\ 
%  \hline
% \end{tabular}
%  %\vspace{-3mm}
% \end{table}

Table~\ref{Table_hierarchical_model} presents the performance of the LHC per level in the hierarchy using the hierarchical classification metrics. For the first level in the hierarchy, the flat classification metric is adopted because no error is propagated into this level (i.e., it is the root node). For level 2 and 3, error from level 1 is propagated into these levels, so this error must be penalized. Hence, the hierarchical classification metrics are used for the two levels. We assessed the LHC when only the 26 features from HHT are utilized to build each classifier in the hierarchy. A precision of 90.23\%, 67.86\%, and 56.86\% is achieved for level 1, 2, and 3, respectively. When the feature set from HHT-WPT is utilized for modeling, the hierarchical metrics increase. The impact of WPT was seen more in the first-level classifier. The increase in hierarchical precision is 4.03\%, 2.98\%, and 2.48\% for level 1, 2, and 3, respectively. The hierarchical precision of the model decreases as we traverse down the hierarchy.
% \begin{table}[t!]
% \setlength{\tabcolsep}{3pt}
% \centering
% \caption{Performance measure of the local hierarchical classifier at different levels in the hierarchy.}
% \label{Table_hierarchical_model}
% \begin{tabular}{|c|c|c|c|}

% \hline
%  Level & Precision (\%) & Recall(\%) & $F_1$-score (\%)\\
 
% \hline
% 1  & 90.23  &	90.60 &	90.42

%  \\ 	

% \hline
% 2  & 67.86 &	67.86  &	67.86

%  \\ 
%  \hline
% 3  & 56.86 &	56.90 &	56.88

%  \\ 

%  \hline
% \end{tabular}
%  %\vspace{-3mm}
% \end{table}

\section{Conclusion} \label{eight}
This work proposed a framework that uses SDAE-LOF and hierarchical classification for UAV detection and identification. The framework exploits the steady-state of RF signals emanating from the communication between the UAV and its flight controller. Because the UAV communication link operates in the same frequency band (i.e., 2.4~GHz) as WiFi and Bluetooth devices, an SDAE-LOF model was used to detect the presence of a UAV or UAV controller signal with an accuracy of 89.52\%. SDAE is used for the compression of signal into a latent representation that is robust to channel noise. Distance measure and the number of nearest neighbors used for estimating the local density of datapoint are the most critical hyperparameters of the LOF algorithm. The effect of these two factors significantly affects the performance of our proposed SDAE-LOF model. It was observed that a high number of nearest neighbors (i.e., greater than 20) improves the detection accuracy at low SNR. The use of either Manhattan or Euclidean distance as the distance measure has comparable performance in predictive inference time and varying SNR. However, both Manhattan and Euclidean distances outperform Cosine distance.

More so, a hierarchical learning framework is used to classify the type of UAS signal, UAV model, and flight mode of the UAV. The framework uses HHT and WPT for feature extraction. The feature set from WPT augments the feature set from HHT to improve the performance of the framework. The impact of using HHT-WPT for feature extraction is significant at the root node of the hierarchy (i.e., identifying the type of UAS signal) and gradually drops as we traverse down the hierarchy.

Our future work will entail improving the performance of the proposed hierarchical learning framework to achieve higher classification accuracy at the leaf node of the hierarchy and developing a system that will detect adversarial signals that can mislead or attack the functionality of the model.

\section{Acknowledgment}\label{ACK}
The authors would like to thank Dr Ismail Guvenc of the department of electrical and computer engineering at the North Carolina State University, Raleigh, North Carolina. We also appreciate Mr. Evan Arnold and Mr. Michael Picinich of the Institute for Transportation and Education (ITRE), Raleigh, North Carolina for their help in carrying out the experiment in this work.
 
%  and Dr. Mihail Sichitiu of the Department of Electrical Engineering, North Carolina State University, for providing some of the UAV controllers used in this study. Also, we thank Dr. Karthik Vasudeva and Mr. Mark Funderburk for helping out with some of the experiments.

\balance
\bibliography{IEEEabrv,reference}
\bibliographystyle{IEEEtran}
% \bibliography{main.bbl}
\end{document}